\documentclass[twoside]{article}

\usepackage{lipsum} 

\usepackage[round]{natbib}
\usepackage[sc]{mathpazo} 
\usepackage[T1]{fontenc} 
\usepackage{microtype} 

\usepackage[hmarginratio=1:1,top=32mm,columnsep=20pt]{geometry} 
\usepackage{multicol} 
\usepackage[hang, small,labelfont=bf,up,textfont=it,up]{caption} 
\usepackage{booktabs} 
\usepackage{float} 
\usepackage{lettrine} 
\usepackage{paralist} 

\usepackage{abstract} 

\usepackage{titlesec} 
\renewcommand\thesection{\Roman{section}} 
\renewcommand\thesubsection{\Roman{subsection}} 
\titleformat{\section}[block]{\large\scshape\centering}{\thesection.}{1em}{} 
\titleformat{\subsection}[block]{\large}{\thesubsection.}{1em}{} 

\usepackage{fancyhdr} 
\usepackage{graphicx}
\DeclareGraphicsExtensions{.jpg}
\usepackage{caption}
\usepackage{subcaption}
\usepackage{amsmath}
\usepackage{booktabs,xcolor,colortbl}
\usepackage{hhline}
\usepackage{color,soul}
\usepackage[hidelinks]{hyperref}
\usepackage{wrapfig}

\title{\vspace{-15mm}\fontsize{24pt}{10pt}\selectfont\textbf{Ketamine-Medetomidine General Anesthesia Occurs With Alternation of Cortical Electrophysiological Activity Between High and Low Complex States
}} 

\author{
\large
\textsc{Eduardo C. Padovani}
\thanks{Email: \texttt{eduardo.padovani@alumni.usp.br}}\\[2mm] 
\normalsize  \\ 
\normalsize 
\vspace{-5mm}
}
\date{}

\begin{document}

\maketitle 

\thispagestyle{fancy} 


\begin{abstract}

\noindent \ Anesthetic agents are known to induce a range of alterations in cortical electrophysiological activity, such as the rise of signature patterns, changes in statistical properties, and altered dynamic behavior of neural records. Plenty of methods can be used to track and monitor these changes, among them complexity metrics demonstrated to have the power to discriminate states involving distinct levels of awareness. There is a consensus that anesthetic drugs can interfere with neural activities at different levels and time scales, being able to induce alterations both locally and in the spatiotemporal patterns established throughout the whole cortex. However, it is still unclear how such changes in the complexity of cortical activity are supposed to occur, and experimental evidence is still needed. For this purpose, we have analyzed an ECoG records database of a Ketamine-Medetomidine anesthetic induction experiment in a non-human primate subject. The MDR-ECoG technique provided records of cortical activity with both high temporal and spatial resolution allied with extensive coverage of the cortical surface. The Permutation Entropy and the Fractal Dimension were employed to evaluate the complexity of the neural time series. It was found that the complexity of cortical activity was relatively constant during awakened conditions. The transition to unconsciousness occurred relatively fast; it required about 30 to 40 seconds for the first remarkable changes to take place. During general anesthesia, the complexity assumed a considerable variation at local levels and fluctuated apparently without a defined period. The cortex dynamically alternated between high and low complex states on a global scale. This study provides novel evidence of the effects of anesthetics on neural activity and cortical dynamics, complementing the actual scenario for elucidating anesthetics mechanisms in terms of circuits, pathways, and global brain functioning.

\end{abstract}


\begin{multicols}{2} 

\section{Introduction}
\lettrine[nindent=0em,lines=3]{E}lectrophysiological cortical records are known to exhibit highly complex patterns, and it is a fact that changes in levels of awareness, whether caused by anesthetic agents, epileptiform activity, or sleep, reflect in signals with characteristic patterns that are distinguished from those typically observed under alert conditions. Since electrophysiological activity signals contain relevant information about the physiological state of individuals, a variety of methods can be used to evaluate changes in behavior, patterns, and the statistical properties of neural time series. 
In particular, when we are interested in making inferences about different levels of awareness, it is possible to highlight complexity-related metrics, which have proven to have the power to discriminate physiological states involving distinct levels of consciousness. Because of this, complexity metrics are even used in clinical medicine to monitor patients during procedures involving sedation and general anesthesia. However, in these applications and neuroscience experiments, most of the time, a reduced number of electrodes are used, thus giving information about changes that occur locally in certain cortical regions without providing detailed information about what is happening in the cortex as a whole.

Since anesthetic agents can concomitantly interfere with local molecular environments, neuronal circuits in the thalamic nuclei, circuits widespread throughout the cortex, and also the communication established between the thalamus and the cortex, anesthetic drugs are able not only to affect the activity of specific cortical regions but in fact can interfere with the dynamics of neural activities in the brain as a whole, at a systems level. In this manner, we assert that the elucidation of the spatiotemporal organization of the complexity of neural activity over extensive cortical areas may bring new relevant information to complement the actual scenario and contribute with novel experimental evidence for the comprehension of the effects of general anesthesia in terms of neural circuits, pathways, and global brain functioning.

The objective of the present research was to verify and describe how statistical properties associated with the complexity of cortical electrophysiological activity change as soon as an anesthetic induction agent is administered to a subject. Particularly verifying the changes in patterns and dynamics that occur along with the unfolding of the transition to unconsciousness, as well as the spatiotemporal organization of the cortical activity’s complexity during general anesthesia.

For this objective, a database respective to an experiment that involved \textit{Ketamine-Medetomidine} anesthetic induction in an old-world monkey was analyzed. In this database, electrophysiological neural activity was recorded using a technique that simultaneously offered extensive coverage of cortical surfaces along with high spatial and temporal resolution. Two different complexity-metrics methods, the \textit{Fractal Dimension} \citep{higuchi1988approach} and the \textit{Permutation Entropy} \citep{bandt2002permutation} were used to infer the complexity of the neural time series.

\vspace{0.9\baselineskip}

In this research, we have characterized the complexity of the electrophysiological neural activity in the awakened resting state, Ketamine-Medetomidine general anesthesia, and during the transition to induced unconsciousness. We have analyzed the complexity of distinct cortical regions at a local level and the spatiotemporal patterns established throughout the cortex at a systems level.  
Under resting-state conditions, the complexity of cortical activity was demonstrated to be relatively constant over time, despite some events of reduced complexity that appeared and vanished quite frequently, taking place mainly over the temporal and occipital lobes.
The transition to unconsciousness was demonstrated to be relatively fast, requiring about 30 to 40 seconds for the first remarkable alterations to occur. First, the complexity decreased in the occipital and temporal lobes and later diminished in the parietal and frontal lobes. However, during the transition, the central sulcus and areas nearby continued to display high complexity.
 We contemplated that administering the anesthetics promoted novel specific dynamics and not a single overall decrease effect on the complexity of brain activity. Locally, the complexity assumed a higher variation and randomly fluctuated without having a definite period.
 During general induced anesthesia, the whole cortical spatiotemporal patterns dynamically alternated among high and low complex states; the states of high complexity observed had some features that resembled those respective to the awakened resting state. Nonetheless, we verified that both conditions were distinguishable at all times.

\section{Methods}

\subsection{Neural Records Database}

In the present study, a database of cortical electrophysiological activity was analyzed. The database came from the Laboratory of Adaptive Intelligence at the Riken Brain Science Institute, Japan. All surgical and experimental procedures were idealized and performed by the researchers of the affiliated institution. The Riken's scientific ethics committee approved experiments following the experimental protocols (No. H24-2-203(4)) and the recommendations of the Weatherall report: "The use of non-human primates in research." Detailed information regarding methodology, subjects, and materials is available at \citep{nagasaka2011multidimensional} and (\texttt{http://neurotycho.org}).
 
 \enlargethispage{1.5\baselineskip}

 The electrophysiological activity data analyzed is respective to an experiment that involved anesthetic induction in a non-human primate subject of the species \textit{Macaca fuscata}. The data recording technique adopted was the MDR-ECoG, which is considered one of the most advanced technologies available. This technique provided extensive coverage of the cortical surface and offered concomitant high spatial ($5\, mm$) and high temporal resolution ($1 \, KHz$). The animal subject had an array of 128 ECoG electrodes chronically implanted in the subdural space covering the lateral cortical surface of the right brain hemisphere as well as some electrodes at the frontal and occipital medial walls.

In the experiment, the macaque was blindfolded and restrained in an appropriate chair. Neural activity was recorded for approximately 10 minutes under these conditions. After that, a cocktail of Ketamine and Medetomidine was administered for the induction of general anesthesia; after that, neural activity continued to be monitored for the next 25 minutes.

\vspace{-2mm}

\subsection{Complexity-Metrics Methods:}

\subsubsection{Permutation Entropy}

For the calculation of the Permutation Entropy, we have used the algorithm presented and described in \citep{bandt2002permutation}. For each of the 128 time series from the ECoG electrode array, the Permutation Entropy was calculated serially over time throughout the experiment as a sliding window. For each calculation, $2000$ points of the time series were used, equivalent to $2$ seconds of recording of neural activity, with the order parameter of the Permutation Entropy $m=4$ and the time delay $t=15$.

\subsubsection{Fractal Dimension}

For each of the 128 time series from the ECoG electrode array, the Fractal Dimension values were calculated serially over time as a sliding window. To decrease the number of measurements, the Fractal Dimension was calculated once every 2.5 seconds, respective to the recording time of the experiment. Higuchi's algorithm \citep{higuchi1988approach} was used to calculate the Fractal Dimension. For each calculation, $1000$ points of the time series were used (equivalent to 1 second of recording the neural activities), with the time interval parameter $k=100$.

\vspace{-2mm}

\subsection{Statistical Analysis Wilcoxon Signed-Rank Test}

In aiming to verify whether the decrease observed in the complexity during general anesthesia compared to the awake resting state conditions was statistically significant, the Wilcoxon signed-rank test was applied. The resulting p-values of the statistical test are shown in \hyperlink{TABLE1}{$Table \cdot 1$} for the Permutation Entropy and \hyperlink{TABLE2}{$Table \cdot 2$} for the Fractal Dimension.

\vspace{-2mm}

\subsection{t-SNE Plots}
\hypertarget{METHODSIV}{}

We have defined the time-resolved vector states as vectors containing as entries the complexity values at a given time of each one of the 128 electrodes, resulting in 128-Dimensional vectors that were estimated serially over time throughout the experiment. For the permutation entropy, the time-resolved vector states were estimated at every 2.0 seconds of the recording experiment and for the Fractal Dimension at every 2.5 seconds.
The time-resolved vector states (128-D) were used as input features into the t-SNE algorithm \citep{van2008visualizing} implemented in the R-CRAN package Rtsne \citep{krijthe2018package}, with the parameters: \textit{perplexity} $=30$, \textit{exaggeration factor} $=12$, and the \textit{maximum number of interactions} $=500$. This analysis was independently performed for the Fractal Dimension and for the Permutation Entropy.

\section{Results}

In this research, we were able to track the spatiotemporal patterns of the cortical electrophysiological activity’s complexity in the macaque under study. The dynamic patterns were characterized in awakened resting-state conditions and Ketamine-Medetomidine general anesthesia. A substantial difference between these two states has been verified. The unfolding dynamics of the loss of consciousness have also been observed, and we were able to follow the alterations that occurred in specific cortical areas as well as in the cortex at a global level. 

Two different methodologies were employed to estimate the complexity of neural time series: the Fractal Dimension and the Permutation Entropy. Although some distinctions were observed among the results from each methodology, in a broad sense, the findings obtained were reasonably equivalent, and we were able to draw the same conclusions from both approaches.

\enlargethispage{1.5\baselineskip}

\subsection{Awakened Resting State}

By analyzing the figures of the Permutation Entropy (\hyperlink{FIGURE1}{$Figure \cdot 1$}) and the Fractal Dimension (\hyperlink{FIGURE6}{$Figure \cdot 6$}) over time throughout the anesthetic induction, we verified that during resting state conditions, the complexity of cortical activity did not present an expressive and remarkable variation, in general being considerably constant over time (\textit{see \hyperlink{FIGURE1}{$Figure \cdot 1$} \textit{and} \hyperlink{FIGURE6}{$Figure \cdot 6$}, the values obtained before the vertical red line at 10.5 minutes in both figures}). Regarding the magnitude of the values, during the resting state, the Permutation Entropy varied in the range of $\approx 3.1$ to 3.2 (see \hyperlink{FIGURE1}{$Figure \cdot 1$}), and the Fractal Dimension in the range of $\approx 1.7$ to 1.8 (see \hyperlink{FIGURE6}{$Figure \cdot 6$}). We also noticed some differences in the dynamic behavior according to the position of the electrodes over the cortex. Among all electrodes, those positioned over the frontal lobe (\textit{electrodes A to L}) were the ones that exhibited the lowest variation, whereas electrodes located over the occipital and temporal lobes (\textit{electrodes N to V}) were more prone to display some variation (\textit{see \hyperlink{FIGURE1}{$Figure \cdot 1$} and \hyperlink{TABLE1}{$Table \cdot 1$} for the Permutation Entropy; \hyperlink{FIGURE6}{$Figure \cdot 6$} \textit{and} \hyperlink{TABLE2}{$Table \cdot 2$} for the Fractal Dimension}).
By analyzing the values of complexity over the coordinates of the electrodes, we verified the spatiotemporal patterns respective to the complexity of neural activity over distinct cortical regions, as well as the behavior of the cortex at a system's level during the awakened resting-state conditions (see \hyperlink{FIGURE3}{$Figure \cdot 3$}, and \hyperlink{FIGURE8}{$Figure \cdot 8$}). We have noticed that the complexity of the cortical activity was approximately the same over the whole cortex; no particular region that stood out for presenting complexity intrinsically significantly higher or inferior all the time was verified. We observe in \hyperlink{FIGURE3}{$Figure \cdot 3$} and also in \hyperlink{FIGURE8}{$Figure \cdot 8$} that the general aspect and most frequent patterns consisted of the majority of the electrodes displaying a predominantly red color according to the gradient chart. Localized events characterized by a reduction in the complexity of cortical activity were quite common, although these consisted of events that constantly appeared and vanished, not being present all the time. These events occurred mainly at the occipital and temporal lobes (\textit{see \hyperlink{FIGURE3}{$Figure \cdot 3$} and \hyperlink{FIGURE8}{$Figure \cdot 8$})}.

\vspace{-2mm}

\subsection{Transition}

By analyzing the Figures of the Permutation Entropy and the Fractal Dimension over time throughout the anesthetic induction (see \hyperlink{FIGURE1}{$Figure \cdot 1$}, and  \hyperlink{FIGURE6}{$Figure \cdot 6$}), it was verified that after the administration of the anesthetics (\textit{indicated by the vertical red line at 10.5 minutes in both Figures}), the complexity of the electrophysiological activity remained the same without presenting noticeable changes for approximately 1.5 to 2.0 minutes.
 After this time interval, we were able to contemplate an abrupt and expressive transition, which seemed to be consistent with a single-step process due to its rapid occurrence. The previously higher and approximately constant values presented a decrease and started to exhibit a considerably more significant variation (\textit{see \hyperlink{FIGURE1}{$Figure \cdot 1$}, $Sub$-$Figures$ A to V; and \hyperlink{FIGURE6}{$Figure \cdot 6$}, $Sub$-$Figures$ A to L}). 

By analyzing the values of complexity over the coordinates of the electrodes, we could contemplate how distinct specialized areas, as well as the cortex at a system's level, behaved along with the transition (\textit{see} \textit{\hyperlink{FIGURE4}{$Figure \cdot 4$}, and \hyperlink{FIGURE9}{$Figure \cdot 9$}}). Changes were noticeable in both measures of complexity evaluated, (\textit{see} \hyperlink{FIGURE4}{$Figure \cdot 4$}) for the Permutation Entropy and (\textit{see} \hyperlink{FIGURE9}{$Figure \cdot 9$})  for the Fractal Dimension. Out of these two pictures, the one concerning the Fractal Dimension (\hyperlink{FIGURE9}{$Figure \cdot 9$}) had the changes in the spatiotemporal patterns quite more apparent.

The first information we can infer from the Figures is the approximate time needed for the changes to occur. In the Permutation Entropy  \hyperlink{FIGURE4}{$Figure \cdot 4$}, we verify that Sub-Figures 8 and 9 presented patterns characteristic of the alert condition, while Sub-Figures 28 and 29 already displayed very distinct patterns from those observed previously (\textit{see} \hyperlink{FIGURE4}{$Figure \cdot 4$}). Since the time interval among consecutive Sub-Figures is 2.0 seconds, it is possible to infer that the change in the state occurred within approximately 30 to 40 seconds (see \hyperlink{FIGURE4}{$Figure \cdot 4$}). Regarding the Fractal Dimension \hyperlink{FIGURE9}{$Figure \cdot 9$}, we verified that Sub-Figures 26 and 27 presented patterns characteristic of the alert condition. In contrast, Sub-Figures 39 and 40 displayed considerably different results from the previous ones (\textit{see} \hyperlink{FIGURE9}{$Figure \cdot 9$}). Considering that the time interval between each Sub-Figure and its subsequent Sub-Figure is 2.5 seconds, we estimate that the change between the patterns occurred within approximately 30 to 40 seconds (\textit{see} $Figure  \cdot  9$). We thus arrive at the same conclusion from both figures.

The unfolding of the transition over different cortical regions can be ascertained by observing the consecutive Sub-Figures of  \hyperlink{FIGURE9}{\mbox{$Figure \cdot 9$}} respective to the Fractal Dimension. First, a substantial decrease in the Fractal Dimension values in the occipital, temporal, and parietal areas occurred without a significant reduction in the electrodes of the frontal region, (\textit{see \hyperlink{FIGURE9}{$Figure \cdot 9$}, $Sub$-$Figures$ $23,\: 24,\: 28,\: 29,\: 31,$ and $32$}). Later on, the complexity of the electrodes located over frontal areas also started to decline (\textit{see \hyperlink{FIGURE9}{$Figure \cdot 9$}, $Sub$-$Figures$ 37, 40, 42, 43, and 48}). A remarkable feature verified along the transition was that electrodes located over the central sulcus and areas nearby showed a tendency to display higher complexity when compared to the rest of the frontal and parietal lobes (\textit{see \hyperlink{FIGURE9}{$Figure \cdot 9$}, $Sub$-$Figures$ 37 to 48}). Another recurrent pattern observed was that both the central sulcus and the occipital lobe displayed a higher complexity in comparison with other areas of the cortex during the transition (\textit{see \hyperlink{FIGURE9}{$Figure \cdot 9$}, $Sub$-$Figures$ 42, 43, 44, 45, and 46}).
All these changes observed in the spatiotemporal patterns contrast with the features found during awake conditions. However, the recurrent spatiotemporal patterns characteristic of general anesthesia that are represented on \hyperlink{FIGURE5}{$Figure \cdot 5$} and \hyperlink{FIGURE10}{$Figure \cdot 10$} only showed up after a few minutes following the occurrence of the first abrupt changes.

\subsection{General Anesthesia}

By analyzing the Figures of the Permutation Entropy (\hyperlink{FIGURE9}{$Figure \cdot 9$}) and the Fractal Dimension ($Figure  \cdot  6$) over time throughout the anesthetic induction, we have noticed that the first remarkable effects on the electrophysiological cortical activity complexity occurred within about 1.5 to 2.0 minutes after the administration of the anesthetics. After a few minutes following these first substantial changes, the dynamics of the complexity of cortical activity assumed a regime that prevailed during general anesthesia.
 It was observed in both \hyperlink{FIGURE1}{$Figure \cdot 1$} and \hyperlink{FIGURE6}{$Figure \cdot 6$} that the cortical activity’s complexity started to assume a wider variation and seemed to fluctuate without possessing a definite period, with the upper limits about the same as the values found during the resting state,  and the magnitude of the lower values was much smaller. Specifically, the Permutation Entropy varied in the range of $\approx$ 2.2 to 3.2, and the Fractal Dimension in the interval of $\approx$ 1.3 to 1.8. By applying the Wilcoxon signed-rank test with a p-value of 5\%, we confirmed that the general decrease observed in the magnitude values of complexity over all the electrodes of the ECoG matrix\footnote{This manuscript does not display the statistical test for all  128 electrodes of the ECoG matrix but displays the most representative ones according to their location over the cortical surface (see \protect\hyperlink{FIGURE12}{Supplementary $Figure \cdot 1$}). } was statistically significant on both Permutation Entropy and Fractal Dimension.

Examining the histograms (\textit{see \hyperlink{FIGURE2}{$Figure \cdot 2$} and \hyperlink{FIGURE7}{$Figure \cdot 7$}}) we can find a considerable distinction among the distributions of the two conditions. During general anesthesia, the distributions tended to be lower and substantially more widespread than during resting-state conditions. This divergence was markedly pronounced in the electrodes located over the frontal lobe.

When we analyze \hyperlink{FIGURE1}{$Figure \cdot 1$} and \hyperlink{FIGURE6}{$Figure \cdot 6$} that comprise the complexity of neural activity over time, we may have the impression that during general anesthesia, the complexity of each electrode fluctuated in a seemingly random manner. However, a remarkable phenomenon was observed when we plotted the complexity values over the coordinates of the electrodes using a color gradient. We have found that, during general anesthesia, the complexity of cortical activity alternates between high and low complex states. There were times when most of the electrodes presented reduced complexity and thus assumed a predominantly blue color. Moreover, there were periods in which most of the electrodes showed high complexity and had mainly red coloring,  apparently resembling the patterns and features observed during alert resting-state conditions (\textit{see \hyperlink{FIGURE5}{$Figure \cdot 5$} and \hyperlink{FIGURE10}{$Figure \cdot 10$}}).

To visualize how the time-resolved complexity vector states (\textit{high-dimensional 128-D, see \hyperlink{METHODSIV}{methods section - IV}}) were spatially distributed, we have used the t-SNE algorithm \citep{van2008visualizing} to project them into a bidimensional map. We have verified that points respective to each experimental condition, awakened resting state, and general anesthesia are positioned at distinct regions of the plane without mixing among themselves (\textit{see \hyperlink{FIGURE11}{$Figure \cdot 11$}}). We have verified these results in both complexity measures used, the Permutation Entropy and the Fractal Dimension. From these findings, we can conclude that, regarding the time-resolved complexity vector states, the awakened resting state and general anesthesia compromise two distinct states that are distinguishable from each other at all times. Although some patterns found during general anesthesia (\textit{see \hyperlink{FIGURE5}{$Figure \cdot 5$} and \hyperlink{FIGURE6}{$Figure \cdot 6$}}) seemed to resemble those found during alert conditions (\textit{see \hyperlink{FIGURE3}{$Figure \cdot 3$} and \hyperlink{FIGURE8}{$Figure \cdot 8$}}), they were in reality distinct.

\section{Conclusions}

In the present research, we inferred the complexity of electrophysiological neural records in a database respective to a Ketamine-Medetomidine anesthetic induction experiment in a macaque subject chronically implanted with a dense ECoG electrode array in the subdural space along the right brain hemisphere. We have inferred the dynamics of the complexity of the neural activity over distinct cortical areas and the spatial-temporal patterns established along the cortex during awakened resting state and general anesthesia conditions. The unfolding of the transition to induced unconsciousness has also been observed. We have verified that during awakened resting-state conditions, the complexity is relatively uniform across the cortex, despite some events of localized complexity reduction frequently occurring at the occipital and temporal lobes. We have found that the first remarkable changes in the complexity of neural activity occurred within about $\approx$ 1.5 to 2.0 minutes after the Ketamine-Medetomidine cocktail was administered and unrolled as a considerably rapid process that took about 30 to 40 seconds to succeed. During general anesthesia, at a local level, the complexity of neural activity assumed a remarkably higher variation, with the maximum values about the same as those found during the resting state and the lower values much smaller. The frontal region was the one that presented the most prominent alterations. Regarding the spatiotemporal patterns assumed by the cortex as a whole, we verified that during Ketamine-Medetomidine general anesthesia, the cortex alternated between high and low complex states. Those highly complex states displayed features that resembled the patterns found during the awakened resting-state conditions. Nonetheless, they were still distinguishable at all times.


\end{multicols}
\begin{figure}[!h]
\begin{subfigure}{.32\textwidth}
  \centering
  \includegraphics[width=1\linewidth]{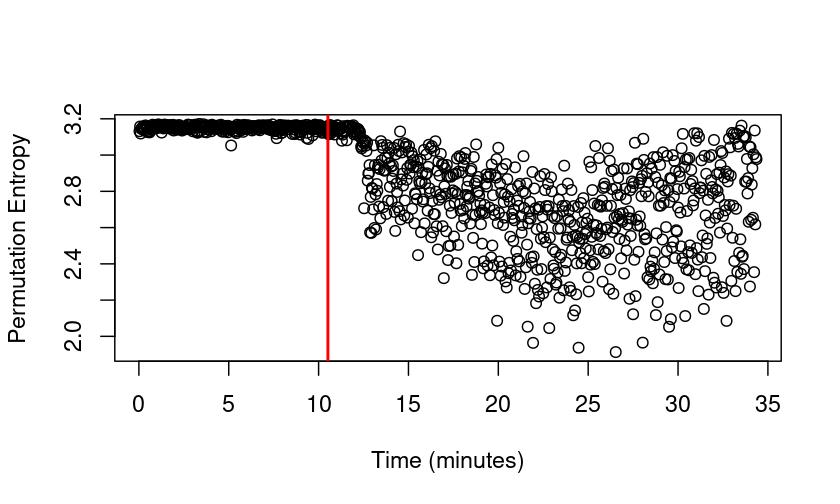}
  \caption{Electrode A}
  \label{fig1:sfigA}
\end{subfigure}%
\begin{subfigure}{.32\textwidth}
  \centering
  \includegraphics[width=1\linewidth]{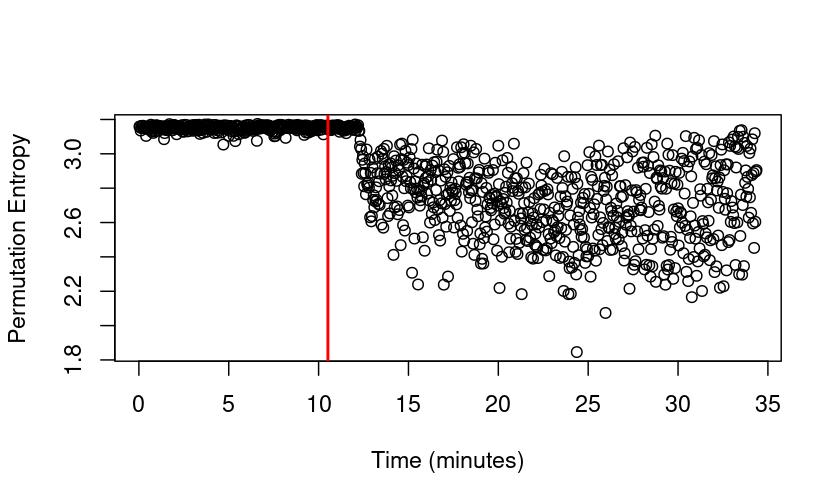}
  \caption{Electrode B}
  \label{fig1:sfigB}
\end{subfigure}%
\begin{subfigure}{.32\textwidth}
  \centering
  \includegraphics[width=1\linewidth]{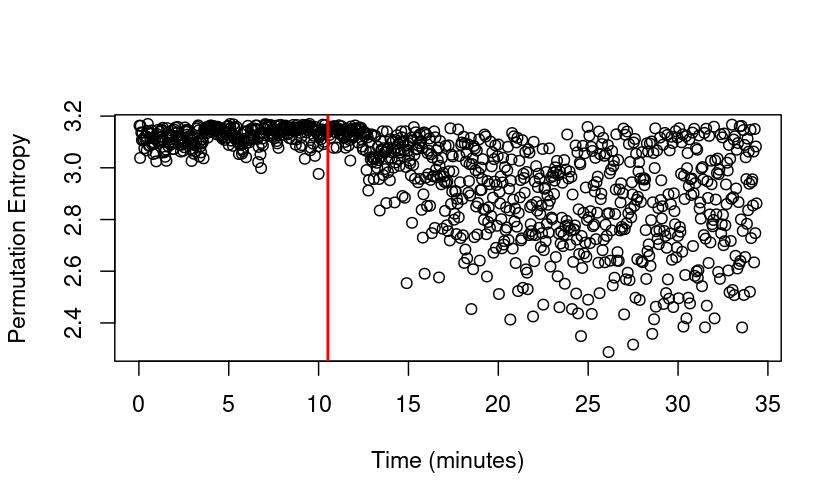}
  \caption{Electrode C}
  \label{fig1:sfigC}
\end{subfigure}\\%
\begin{subfigure}{.32\textwidth}
  \centering
  \includegraphics[width=1\linewidth]{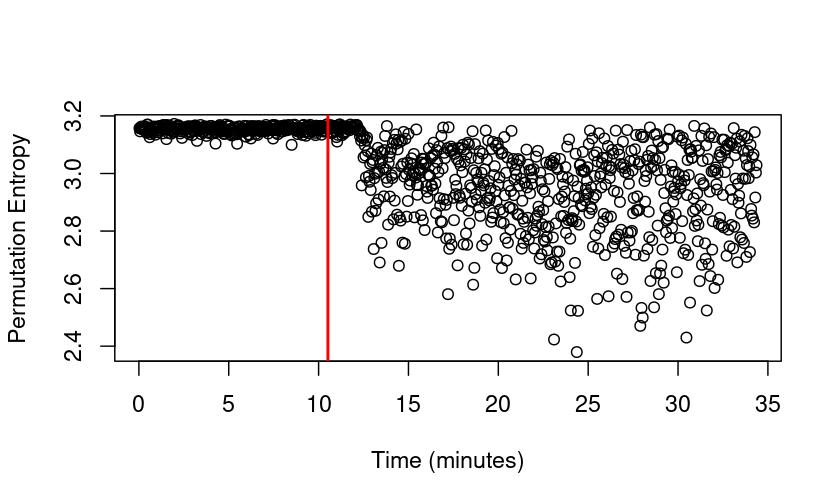}
  \caption{Electrode D}
  \label{fig1:sfigD}
\end{subfigure}%
\begin{subfigure}{.32\textwidth}
  \centering
  \includegraphics[width=1\linewidth]{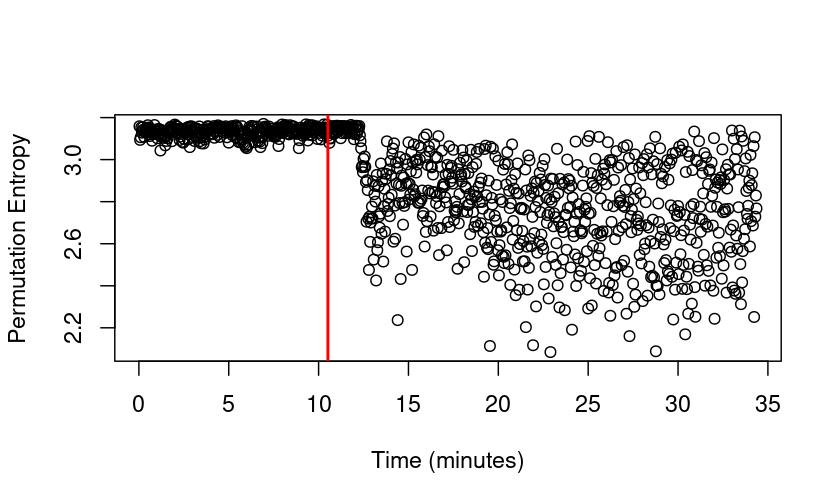}
  \caption{Electrode E}
  \label{fig1:sfigE}
\end{subfigure}%
\begin{subfigure}{.32\textwidth}
  \centering
  \includegraphics[width=1\linewidth]{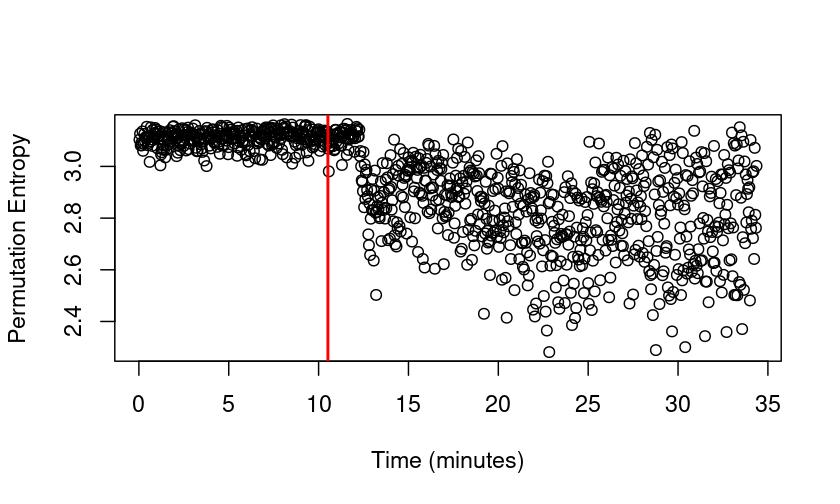}
  \caption{Electrode F}
  \label{fig1:sfigF}
\end{subfigure}\\%
\begin{subfigure}{.32\textwidth}
  \centering
  \includegraphics[width=1\linewidth]{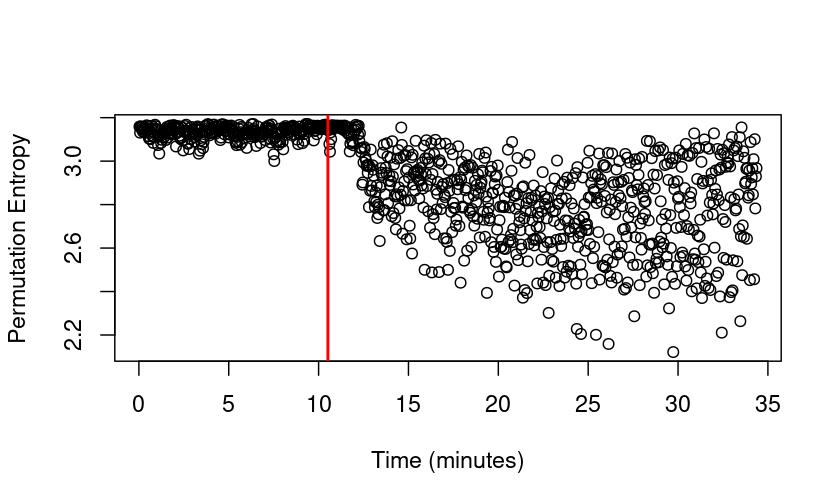}
  \caption{Electrode G}
  \label{fig1:sfigG}
\end{subfigure}%
\begin{subfigure}{.32\textwidth}
  \centering
  \includegraphics[width=1\linewidth]{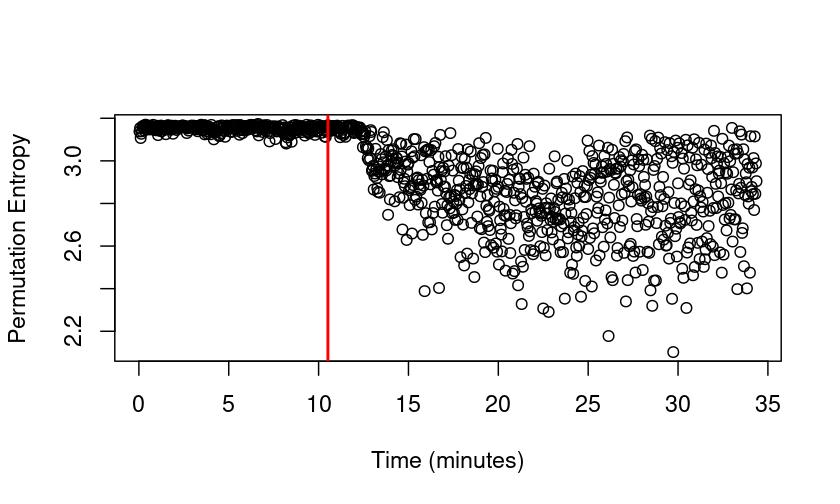}
  \caption{\hypertarget{FIGURE1}{Electrode H}}
  \label{fig1:sfigH}
\end{subfigure}%
\begin{subfigure}{.32\textwidth}
  \centering
  \includegraphics[width=1\linewidth]{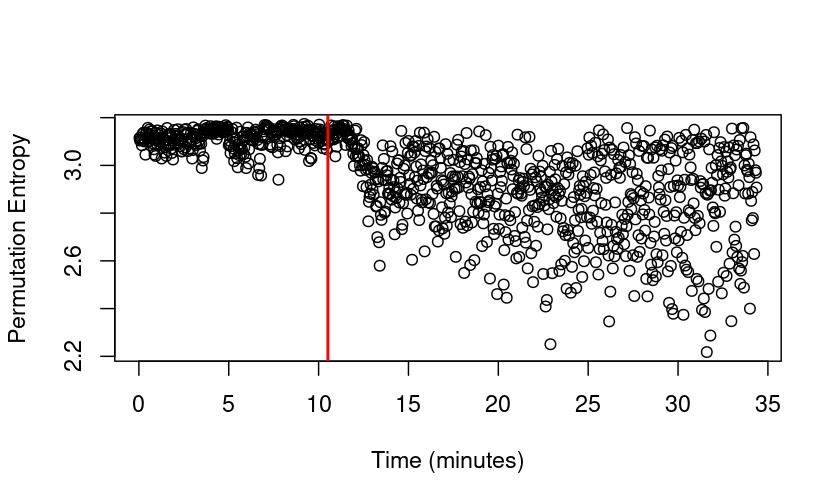}
  \caption{Electrode I}
  \label{fig1:sfigI}
\end{subfigure}\\%
\begin{subfigure}{.32\textwidth}
  \centering
  \includegraphics[width=1\linewidth]{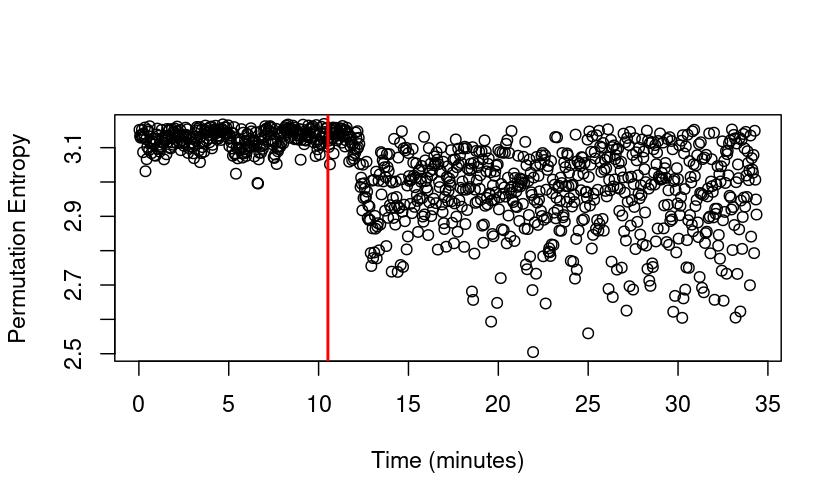}
  \caption{Electrode J}
  \label{fig1:sfigJ}
\end{subfigure}%
\begin{subfigure}{.32\textwidth}
  \centering
  \includegraphics[width=1\linewidth]{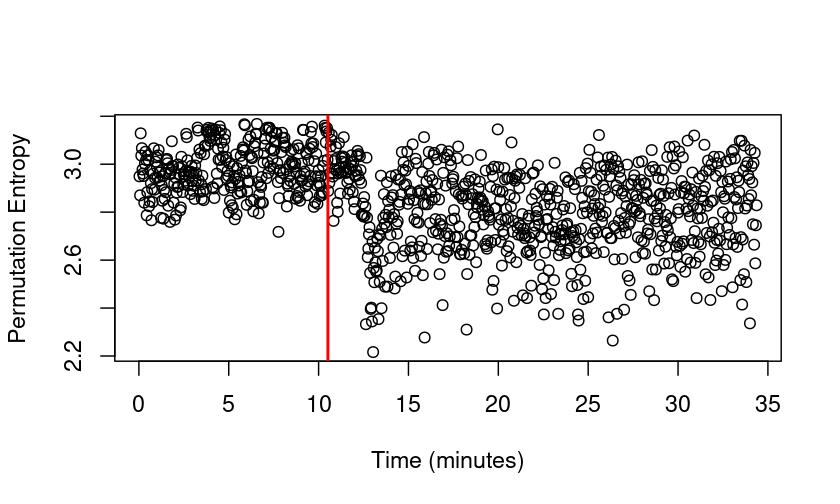}
  \caption{Electrode K}
  \label{fig1:sfigK}
\end{subfigure}%
\begin{subfigure}{.32\textwidth}
  \centering
  \includegraphics[width=1\linewidth]{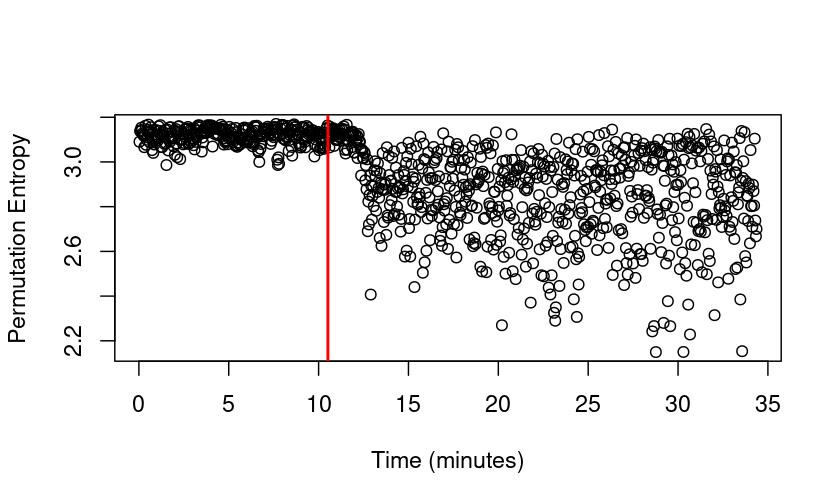}
  \caption{Electrode L}
  \label{fig1:sfigL}
\end{subfigure}\\%
\begin{subfigure}{.32\textwidth}
  \centering
  \includegraphics[width=1\linewidth]{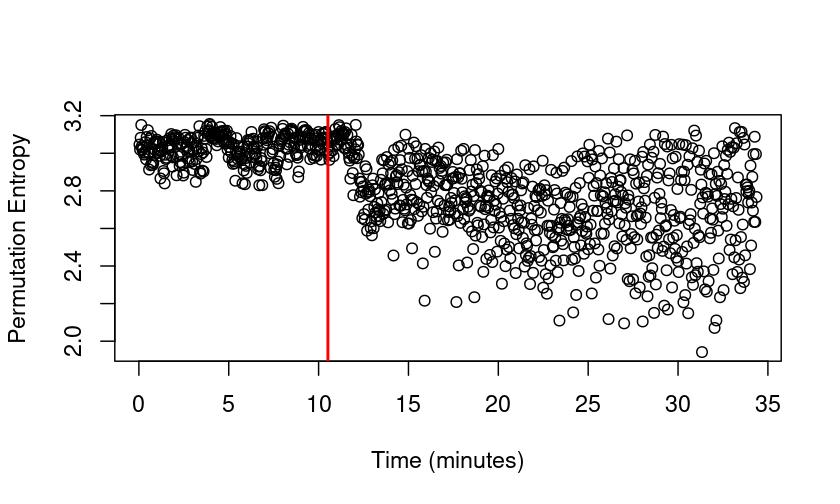}
  \caption{Electrode M}
  \label{fig1:sfigM}
\end{subfigure}%
\begin{subfigure}{.32\textwidth}
  \centering
  \includegraphics[width=1\linewidth]{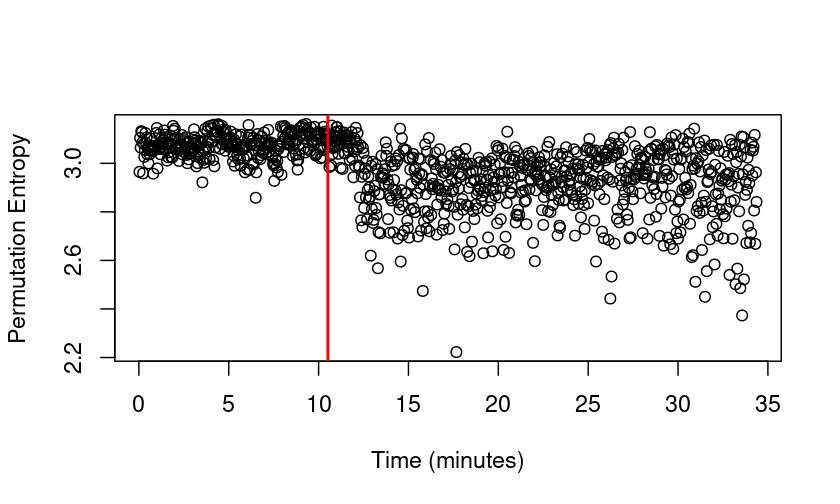}
  \caption{Electrode N}
  \label{fig1:sfigN}
\end{subfigure}%
\begin{subfigure}{.32\textwidth}
  \centering
  \includegraphics[width=1\linewidth]{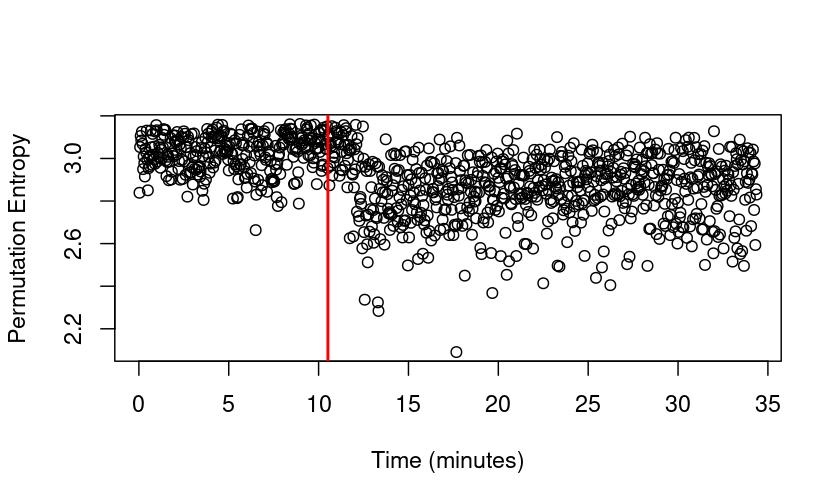}
  \caption{Electrode O}
  \label{fig1:sfigO}
\end{subfigure}%
\caption{\textbf{Permutation Entropy of the electrophysiological time series over time along with the anesthetic induction experiment.}}
\label{figure1A}
\end{figure}
\begin{multicols}{2}


\end{multicols}
\begin{figure}[!h]\ContinuedFloat
\begin{subfigure}{.32\textwidth}
  \centering
  \includegraphics[width=1\linewidth]{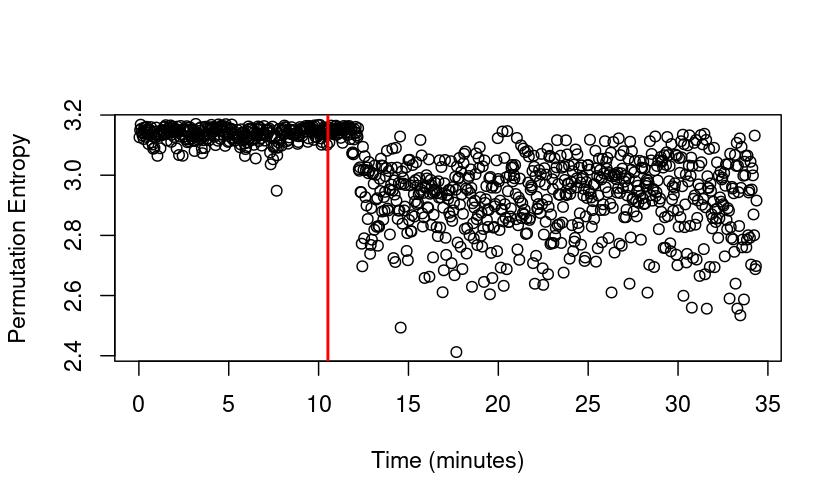}
  \caption{Electrode P}
  \label{fig1:sfigP}
\end{subfigure}%
\begin{subfigure}{.32\textwidth}
  \centering
  \includegraphics[width=1\linewidth]{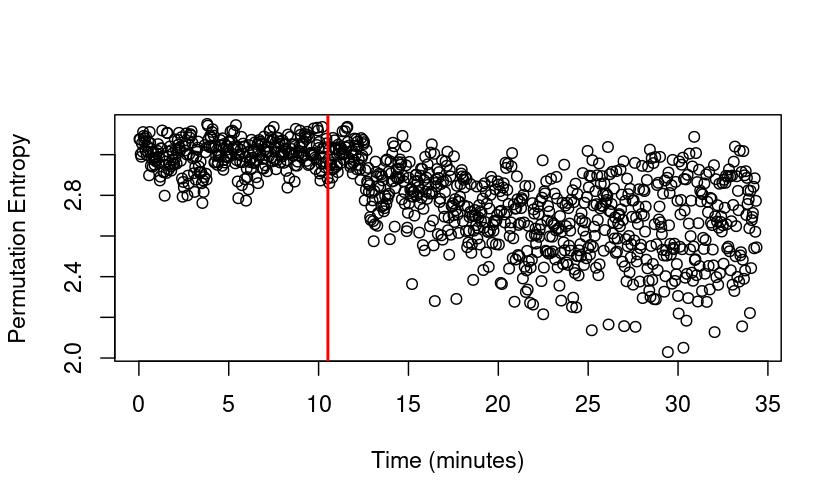}
  \caption{Electrode Q}
  \label{fig1:sfigQ}
\end{subfigure}%
\begin{subfigure}{.32\textwidth}
  \centering
  \includegraphics[width=1\linewidth]{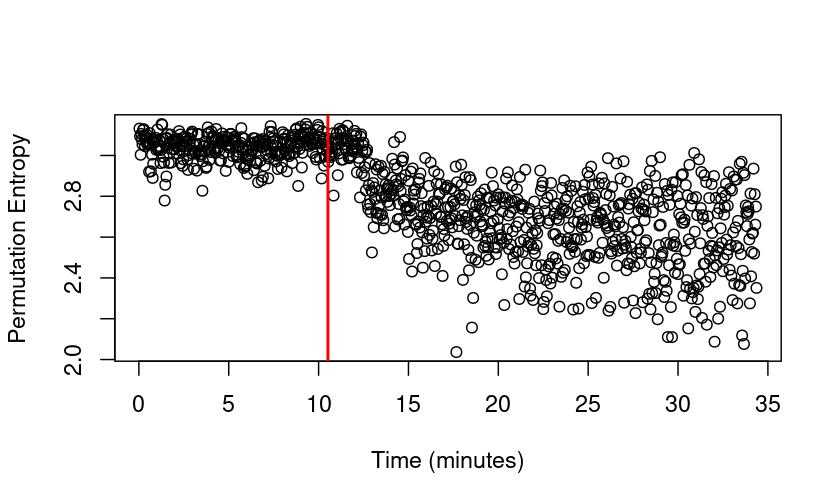}
  \caption{Electrode R}
  \label{fig1:sfigR}
\end{subfigure}\\%
\begin{subfigure}{.32\textwidth}
  \centering
  \includegraphics[width=1\linewidth]{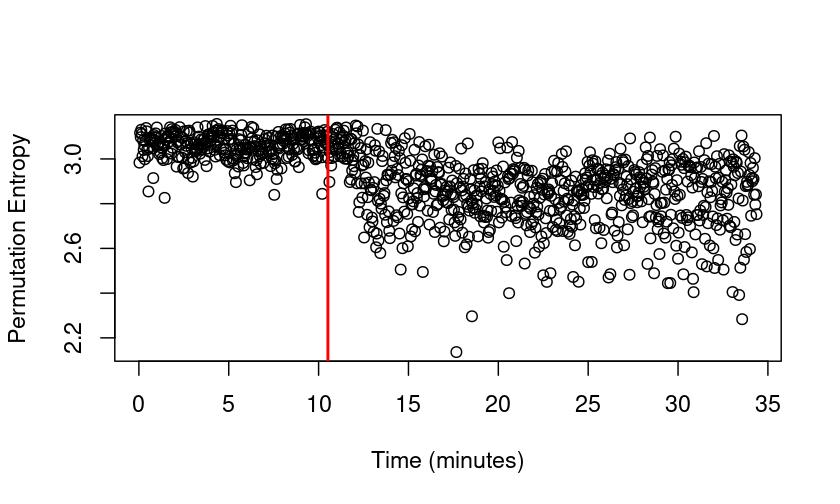}
  \caption{Electrode S}
  \label{fig1:sfigS}
\end{subfigure}%
\begin{subfigure}{.32\textwidth}
  \centering
  \includegraphics[width=1\linewidth]{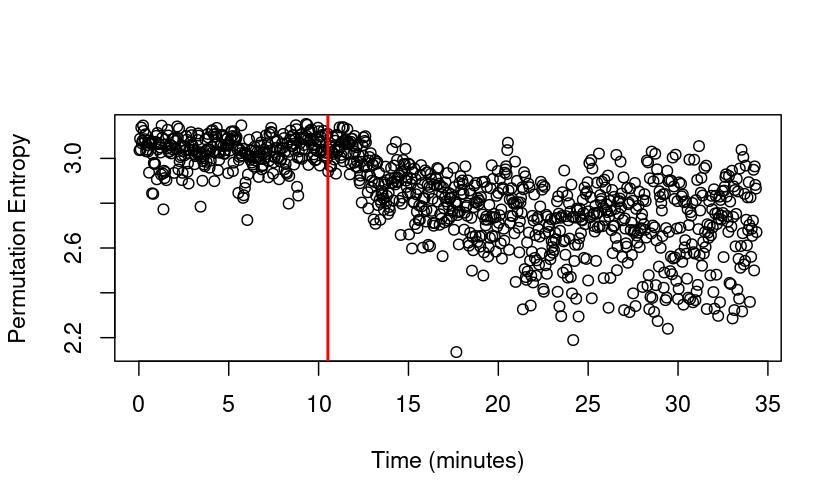}
  \caption{Electrode T}
  \label{fig1:sfigT}
\end{subfigure}%
\begin{subfigure}{.32\textwidth}
  \centering
  \includegraphics[width=1\linewidth]{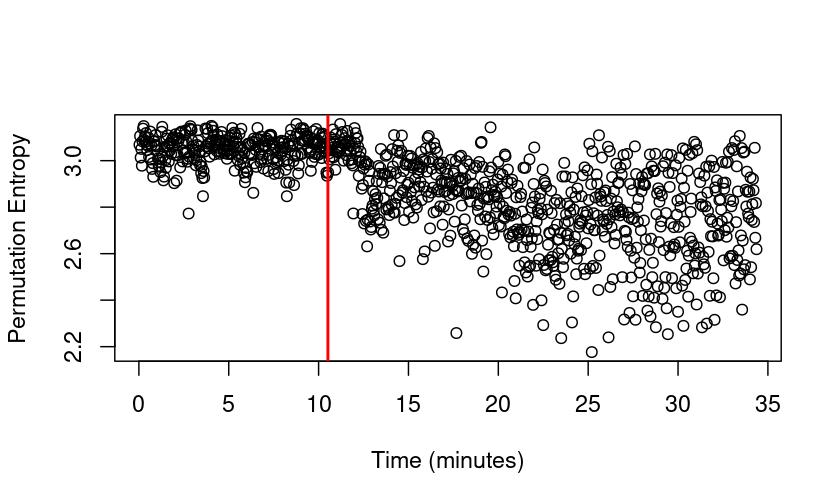}
  \caption{Electrode U}
  \label{fig1:sfigU}
\end{subfigure}\\%
\begin{subfigure}{.32\textwidth}
  \centering
  \includegraphics[width=1\linewidth]{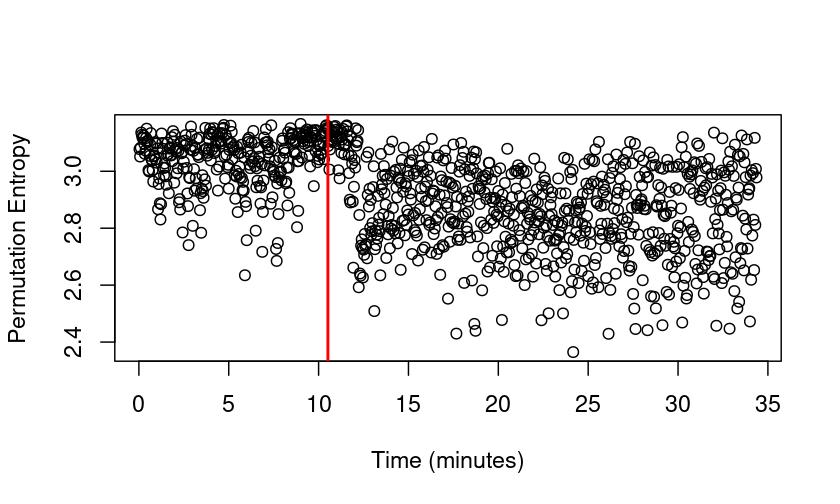}
  \caption{Electrode V}
  \label{fig1:sfigV}
\end{subfigure}%
\caption{\textbf{Permutation Entropy of the electrophysiological time series over the time throughout the anesthetic induction experiment.} Each Sub-Figure is respective to an electrode that was positioned over a specific cortical area, for the exact location of each electrode (see \protect\hyperlink{FIGURE12}{Supplementary $Figure \cdot 1$}, the corresponding letters from A to V). Brain activity started to be recorded in awakening resting-state conditions. At 10.5 minutes the anesthetic drugs were administrated, in each Sub-Figure this event is represented by a vertical red line. Permutation Entropy was calculated at every 2.0 seconds throughout the experiment, being this the time interval between each point and its subsequent one. In general terms, it was noted that in alert conditions, on the majority of the electrodes, the Permutation Entropy of the neural records presented relatively small variation over time. After the administration of the cocktail of anesthetics, the Permutation Entropy remained about the same for about 1.5 to 2.0 minutes, then abrupt changes were verified. It was noticed that the complexity of the neural records assumed a considerably greater variation, being the upper values about the same as those found during alert conditions, and the bottom values were fairly lower. It was also possible to observe that not all electrodes behaved the same, evidencing that the effects of the anesthetics on electrophysiological activities present some variation throughout the cortex.
 }
\label{figure1B}
\end{figure}
\begin{multicols}{2}


\end{multicols}
\begin{table}[!h]

\centering
\caption{\textbf{Mean and standard deviation of the Permutation Entropy values in alert and general anesthesia.} This table presents the mean and standard deviation of the Permutation Entropy values in alert and general anesthesia conditions from the electrodes in distinct cortical regions. For the location of each electrode (see \protect\hyperlink{FIGURE12}{Supplementary $Figure \cdot 1$}, the corresponding letters from A to V). The p-values of the Wilcoxon signed-rank test are shown in the condition that the Permutation Entropy values found during anesthesia are smaller than those observed during the alert resting state. It was verified for all the electrodes that the average of the Permutation Entropy during alert conditions is higher than the average found in general anesthesia and that the standard deviation is smaller during alertness than during anesthesia. Furthermore, for all electrodes, the Wilcoxon test at a p-value of $5\%$ confirmed that the Permutation Entropy of the cortical records decreased during the Ketamine-Medetomidine-induced general anesthesia.
}
\vspace{0.5cm}
\begin{tabular}{l|lr|lr|c}
 \hline
\textbf{Permutation Entropy} & \multicolumn{2}{c}{Awake} \vline & \multicolumn{2}{c}{Anesthesia} \vline & \multicolumn{1}{c}{Wilcoxon Test}\\
\hline
Electrode: & Mean & SD  & Mean & SD & P-Value [Anesthesia $<$ Awake]\\ 
\hline  
\hline
\rowcolor{gray!20}Electrode  A & 3.15 & 0.015 & 2.68 & 0.251 & 4.7e-130  \\
 Electrode  B & 3.15 & 0.017 & 2.70 & 0.224 & 7.9e-131  \\
\rowcolor{gray!20} Electrode  C & 3.12 & 0.036 & 2.87 & 0.196 & 7.8e-95  \\
Electrode  D & 3.15 & 0.012 & 2.93 & 0.149 & 4.1e-124  \\
\rowcolor{gray!20}Electrode  E & 3.13 & 0.025 & 2.75 & 0.223 & 1.3e-127  \\
Electrode  F & 3.11 & 0.034 & 2.81 & 0.176 & 2e-121  \\
\rowcolor{gray!20}Electrode  G & 3.13 & 0.029 & 2.77 & 0.201 & 8.3e-128  \\
Electrode  H & 3.15 & 0.015 & 2.81 & 0.188 & 4.2e-130  \\
\rowcolor{gray!20} Electrode  I & 3.11 & 0.043 & 2.86 & 0.188 & 5.4e-104  \\
 Electrode  J & 3.12 & 0.029 & 2.97 & 0.121 & 2.3e-99  \\
\rowcolor{gray!20} Electrode  K & 2.98 & 0.101 & 2.79 & 0.166 & 2.2e-58  \\
 Electrode  L & 3.12 & 0.037 & 2.84 & 0.198 & 1.3e-115  \\
\rowcolor{gray!20}Electrode  M & 3.03 & 0.073 & 2.70 & 0.224 & 2.6e-99  \\
Electrode  N & 3.07 & 0.050 & 2.91 & 0.134 & 1e-86  \\
\rowcolor{gray!20} Electrode  O & 3.02 & 0.090 & 2.86 & 0.146 & 4e-64  \\
Electrode  P & 3.14 & 0.026 & 2.92 & 0.125 & 5.4e-123  \\
\rowcolor{gray!20}Electrode  Q & 3.00 & 0.075 & 2.68 & 0.199 & 2.9e-110  \\
 Electrode  R & 3.04 & 0.062 & 2.64 & 0.189 & 6.6e-128  \\
\rowcolor{gray!20} Electrode  S & 3.06 & 0.057 & 2.83 & 0.145 & 1.9e-109  \\
Electrode  T & 3.03 & 0.072 & 2.72 & 0.173 & 3.6e-117  \\
\rowcolor{gray!20}Electrode  U & 3.05 & 0.062 & 2.77 & 0.189 & 1.4e-103  \\
Electrode  V & 3.04 & 0.096 & 2.84 & 0.152 & 1.9e-71  \\
\hline
\end{tabular}
\hypertarget{TABLE1}{}
\end{table}
\begin{multicols}{2}


\end{multicols}
\begin{figure}[h]
\begin{subfigure}{.245\textwidth}
  \centering
  \includegraphics[width=1\linewidth]{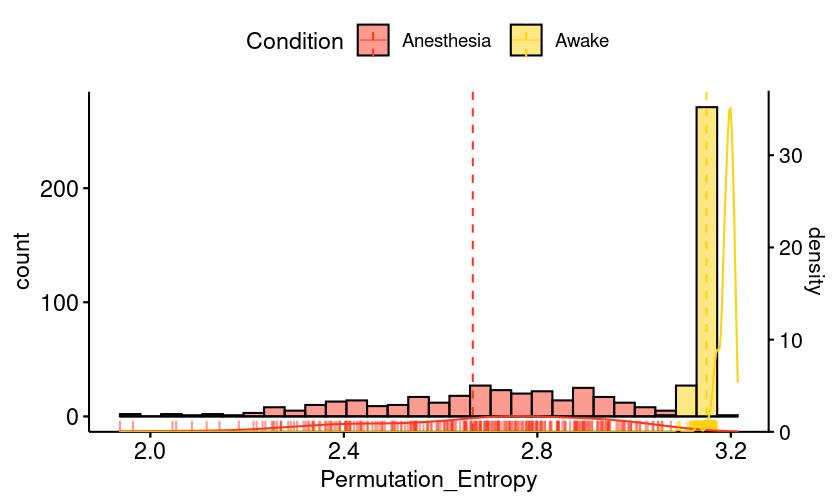}
  \caption{Electrode A}
  \label{fig2:sfigA}
\end{subfigure}%
\begin{subfigure}{.245\textwidth}
  \centering
  \includegraphics[width=1\linewidth]{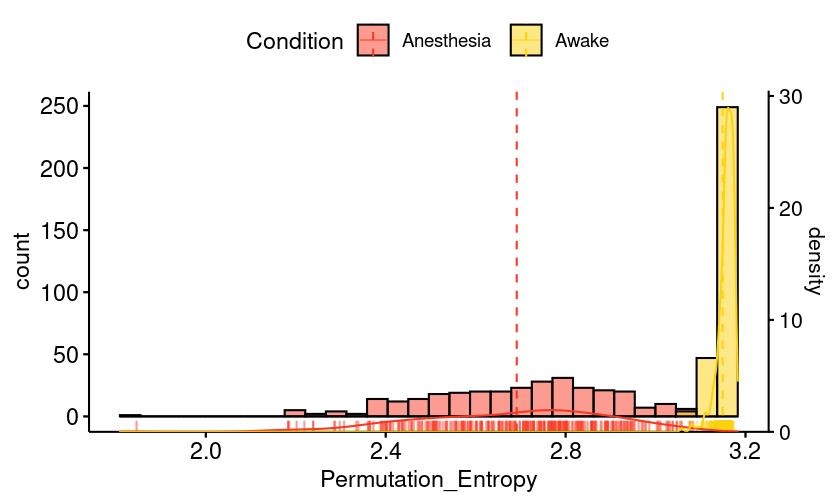}
  \caption{Electrode B}
  \label{fig2:sfigB}
\end{subfigure}%
\begin{subfigure}{.245\textwidth}
  \centering
  \includegraphics[width=1\linewidth]{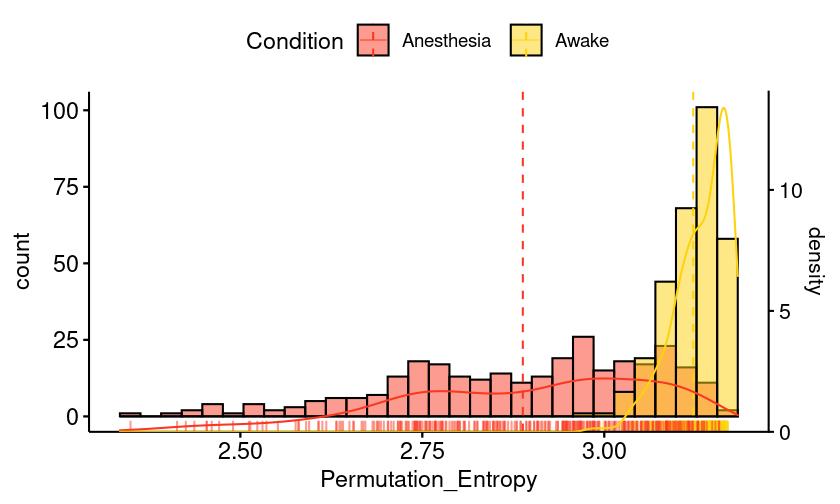}
  \caption{Electrode C}
  \label{fig2:sfigC}
\end{subfigure}%
\begin{subfigure}{.245\textwidth}
  \centering
  \includegraphics[width=1\linewidth]{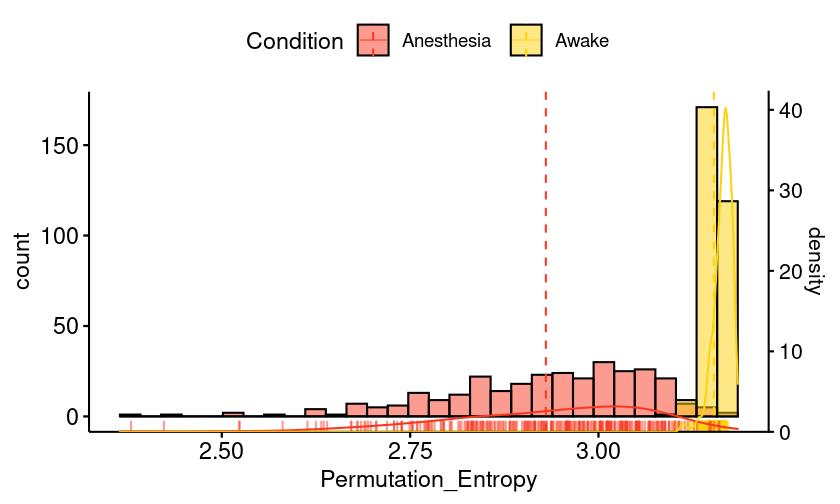}
  \caption{Electrode D}
  \label{fig2:sfigD}
\end{subfigure}\\%
\begin{subfigure}{.245\textwidth}
  \centering
  \includegraphics[width=1\linewidth]{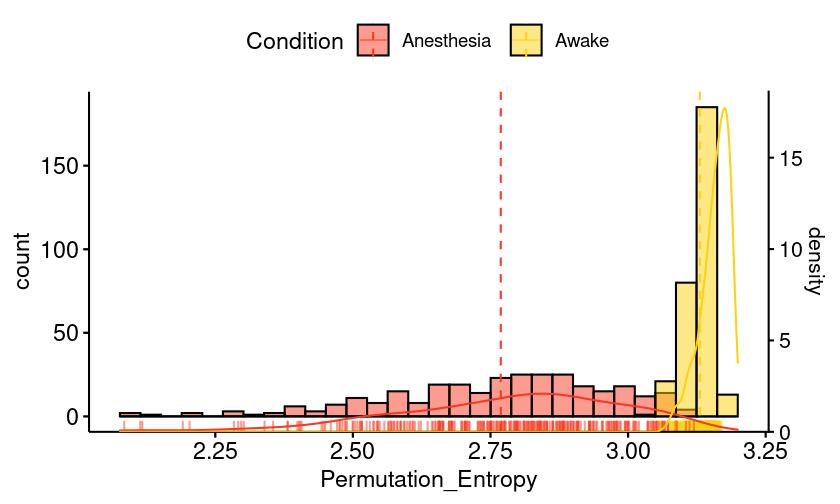}
  \caption{Electrode E}
  \label{fig2:sfigE}
\end{subfigure}%
\begin{subfigure}{.245\textwidth}
  \centering
  \includegraphics[width=1\linewidth]{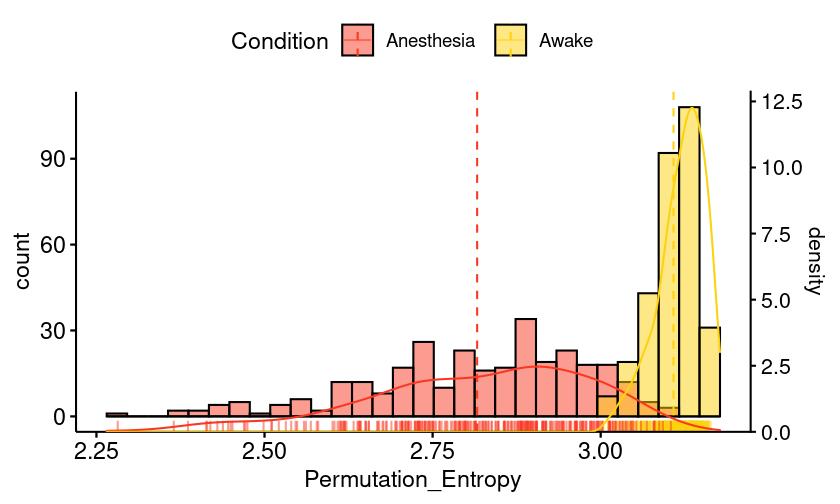}
  \caption{Electrode F}
  \label{fig2:sfigF}
\end{subfigure}%
\begin{subfigure}{.245\textwidth}
  \centering
  \includegraphics[width=1\linewidth]{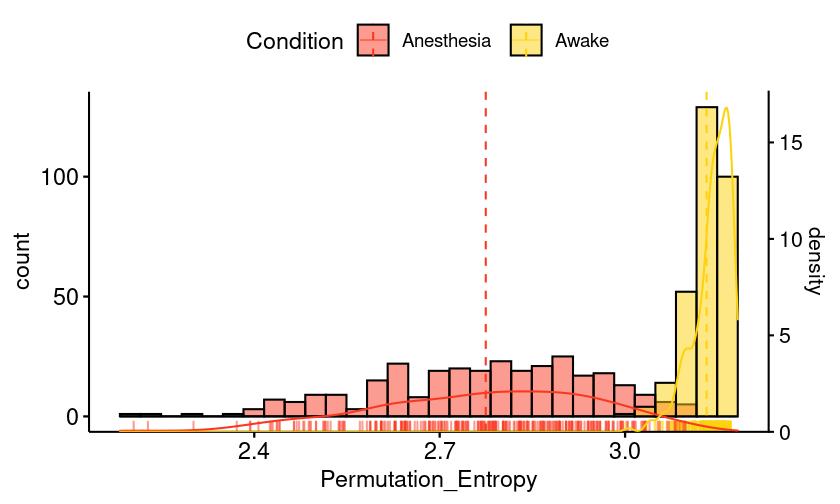}
  \caption{\hypertarget{FIGURE2}{Electrode G}}
  \label{fig2:sfigG}
\end{subfigure}%
\begin{subfigure}{.245\textwidth}
  \centering
  \includegraphics[width=1\linewidth]{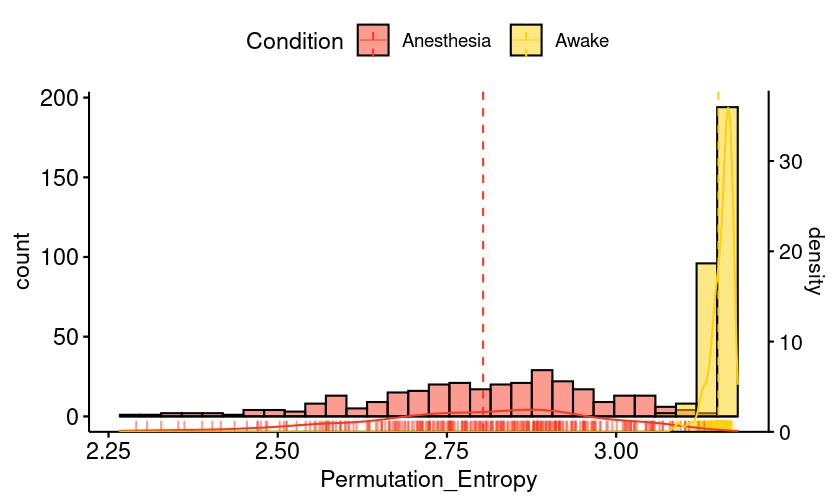}
  \caption{Electrode H}
  \label{fig2:sfigH}
\end{subfigure}\\%
\begin{subfigure}{.245\textwidth}
  \centering
  \includegraphics[width=1\linewidth]{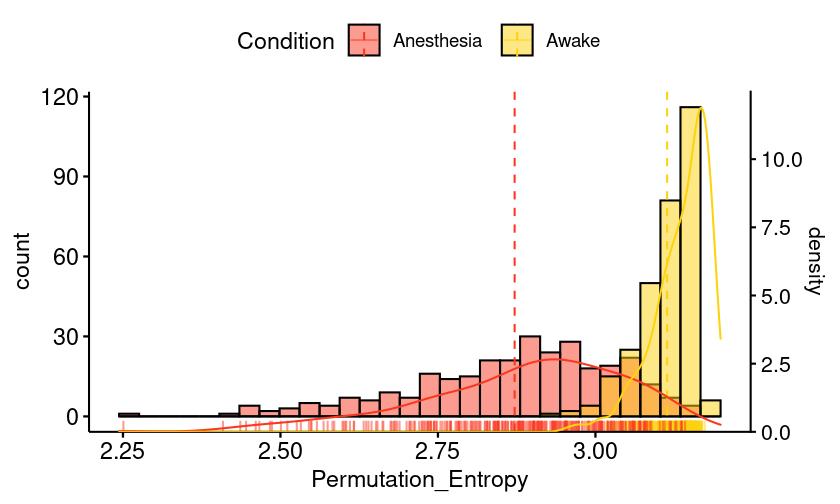}
  \caption{Electrode I}
  \label{fig2:sfigI}
\end{subfigure}%
\begin{subfigure}{.245\textwidth}
  \centering
  \includegraphics[width=1\linewidth]{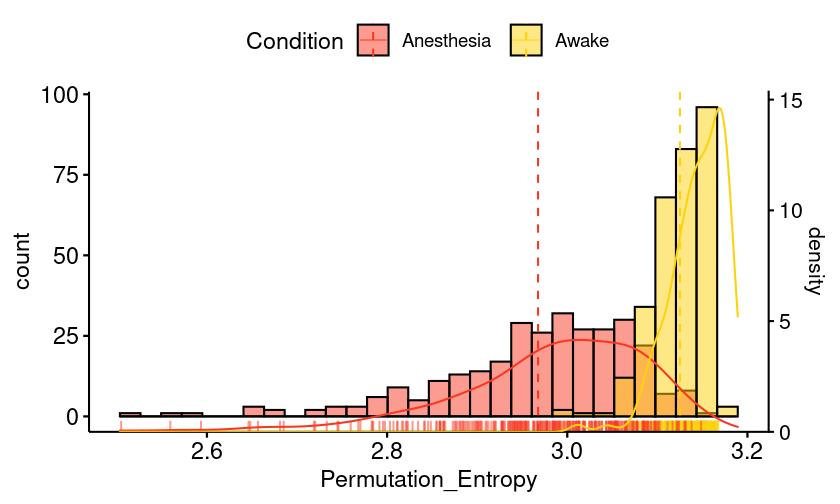}
  \caption{Electrode J}
  \label{fig2:sfigJ}
\end{subfigure}%
\begin{subfigure}{.245\textwidth}
  \centering
  \includegraphics[width=1\linewidth]{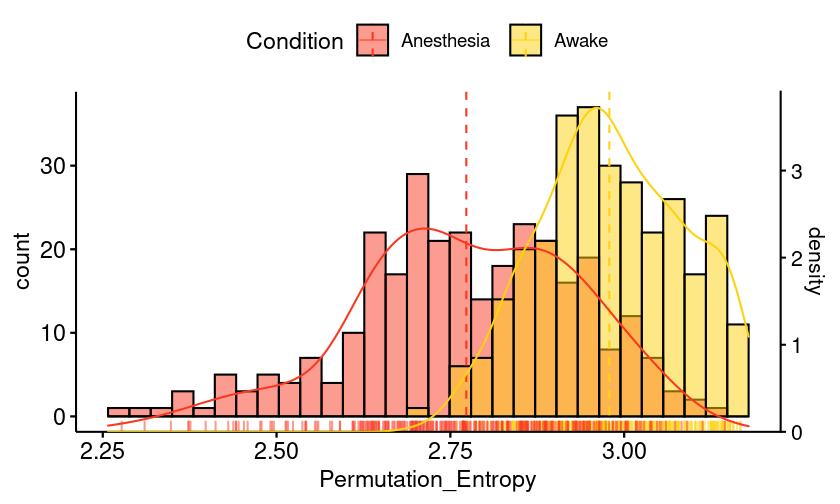}
  \caption{Electrode K}
  \label{fig2:sfigK}
\end{subfigure}%
\begin{subfigure}{.245\textwidth}
  \centering
  \includegraphics[width=1\linewidth]{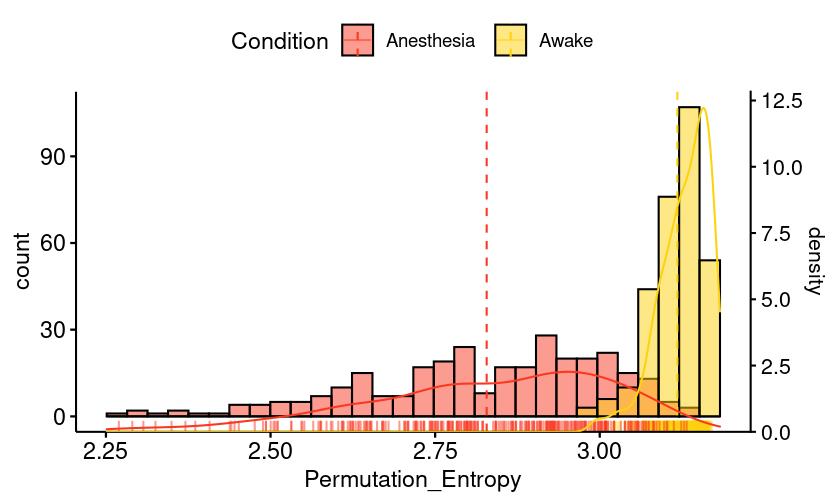}
  \caption{Electrode L}
  \label{fig2:sfigL}
\end{subfigure}\\%
\begin{subfigure}{.245\textwidth}
  \centering
  \includegraphics[width=1\linewidth]{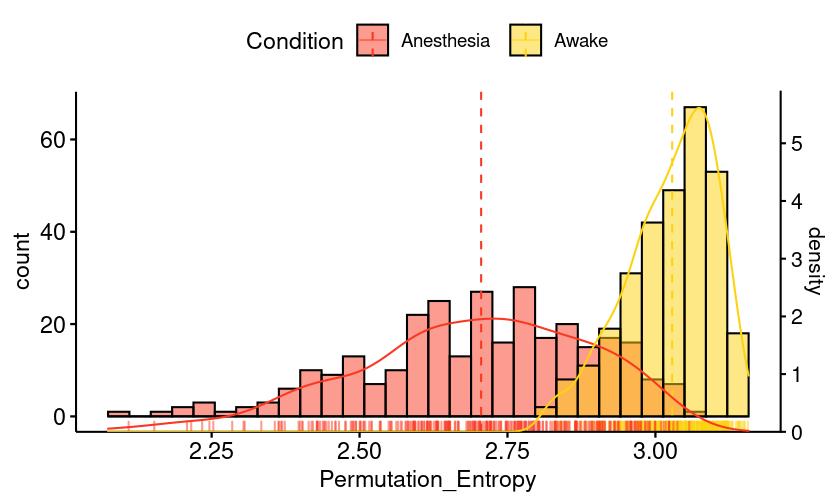}
  \caption{Electrode M}
  \label{fig2:sfigM}
\end{subfigure}%
\begin{subfigure}{.245\textwidth}
  \centering
  \includegraphics[width=1\linewidth]{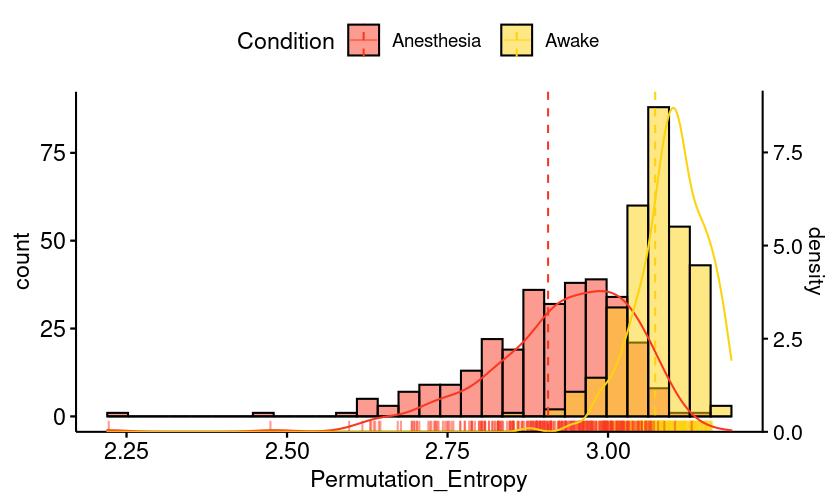}
  \caption{Electrode N}
  \label{fig2:sfigN}
\end{subfigure}%
\begin{subfigure}{.245\textwidth}
  \centering
  \includegraphics[width=1\linewidth]{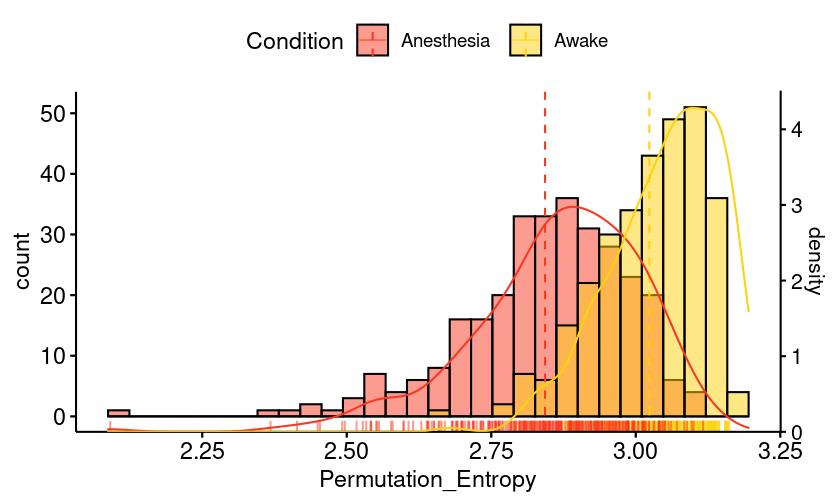}
  \caption{Electrode O}
  \label{fig2:sfigO}
\end{subfigure}%
\begin{subfigure}{.245\textwidth}
  \centering
  \includegraphics[width=1\linewidth]{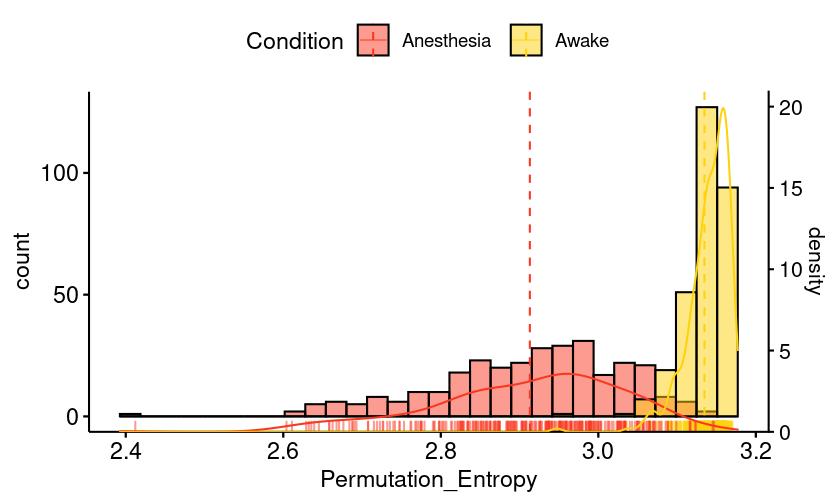}
  \caption{Electrode P}
  \label{fig2:sfigP}
\end{subfigure}\\%
\begin{subfigure}{.245\textwidth}
  \centering
  \includegraphics[width=1\linewidth]{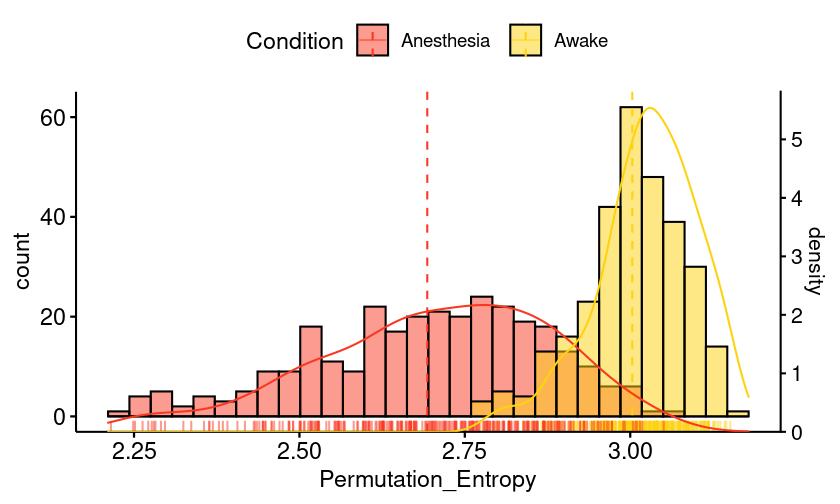}
  \caption{Electrode Q}
  \label{fig2:sfigQ}
\end{subfigure}%
\begin{subfigure}{.245\textwidth}
  \centering
  \includegraphics[width=1\linewidth]{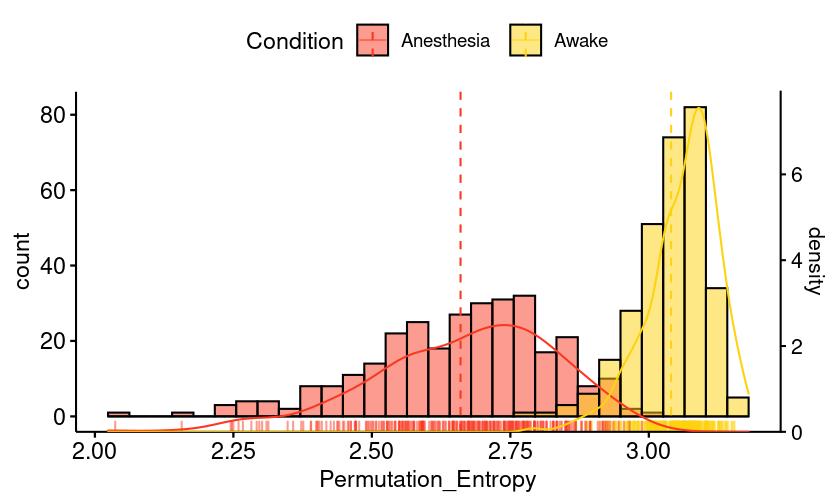}
  \caption{Electrode R}
  \label{fig2:sfigR}
\end{subfigure}%
\begin{subfigure}{.245\textwidth}
  \centering
  \includegraphics[width=1\linewidth]{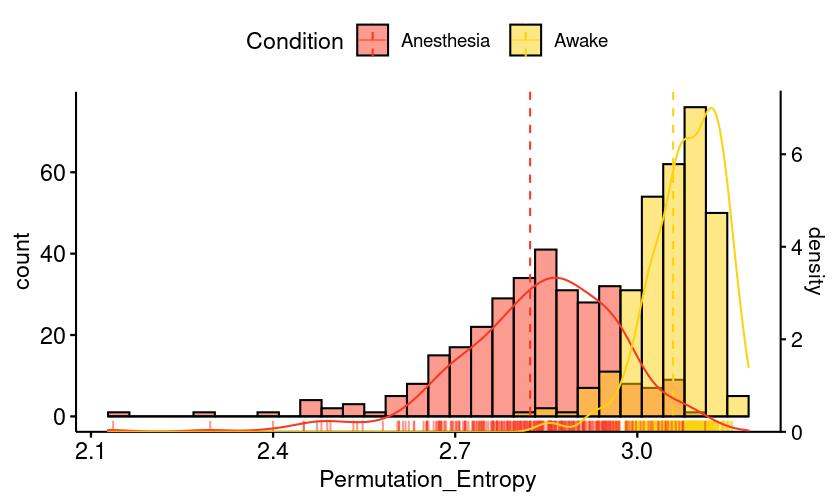}
  \caption{Electrode S}
  \label{fig2:sfigS}
\end{subfigure}%
\begin{subfigure}{.245\textwidth}
  \centering
  \includegraphics[width=1\linewidth]{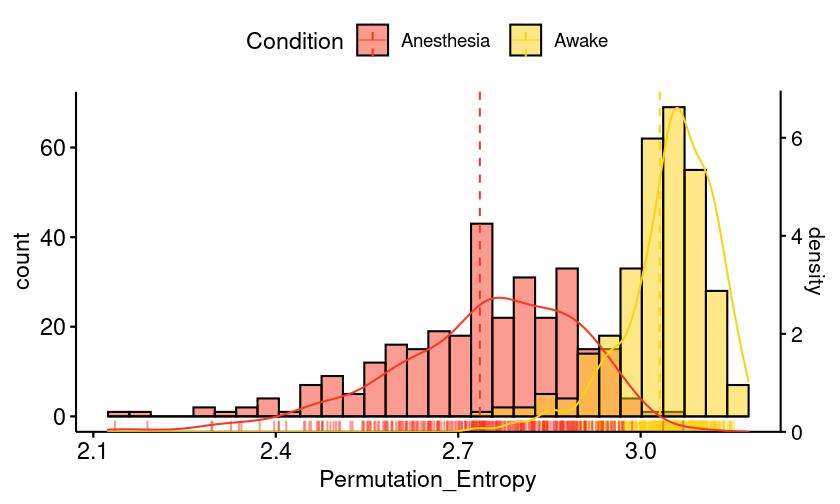}
  \caption{Electrode T}
  \label{fig2:sfigT}
\end{subfigure}\\%
\begin{subfigure}{.245\textwidth}
  \centering
  \includegraphics[width=1\linewidth]{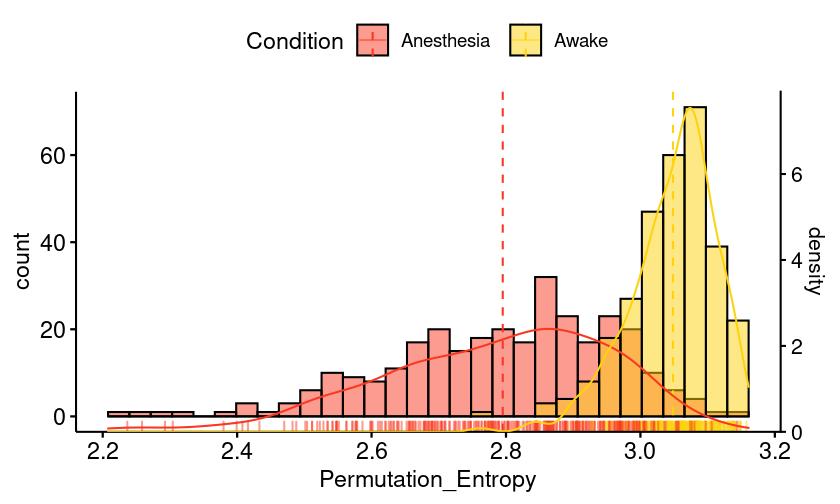}
  \caption{Electrode U}
  \label{fig2:sfigU}
\end{subfigure}%
\begin{subfigure}{.245\textwidth}
  \centering
  \includegraphics[width=1\linewidth]{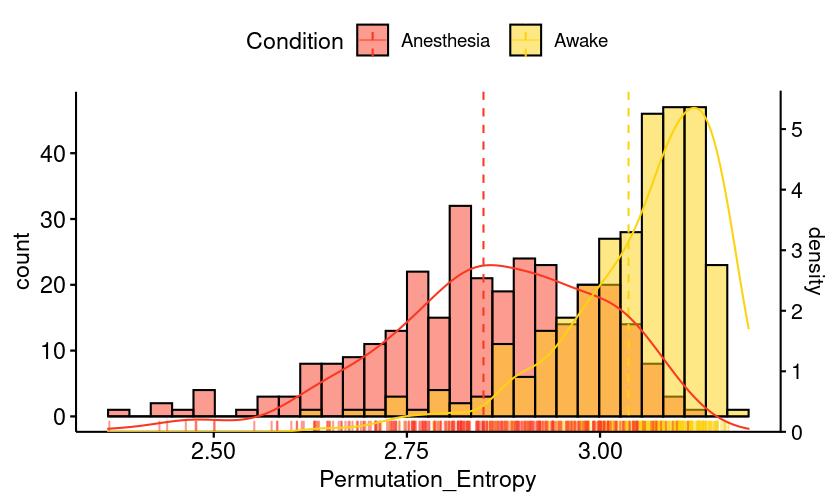}
  \caption{Electrode V}
  \label{fig2:sfigV}
\end{subfigure}%
\caption{\textbf{Overlapping histogram plots of the Permutation Entropy.} Overlapping histogram plots of the electrophysiological time series’s Permutation Entropy respective to the awakened (indicated in yellow) and general anesthesia conditions (indicated in red); see color chart. Sub-Figures correspond to electrodes positioned over a specific cortical region, for the location of each electrode (see \protect\hyperlink{FIGURE12}{Supplementary $Figure \cdot 1$}, the corresponding letters from A to V). 
In general lines, the Permutation Entropy values of the macaque’s cortical electrophysiological activity under general anesthesia tended to be smaller and more widespread when compared to awaken conditions, despite some variations being noticeable over distinct cortical areas. Considering the anesthetic induction, the ECoG electrodes that exhibited the most prominent changes were those positioned over the frontoparietal regions (electrodes A to L, see Sub-Figures A to L) and the electrodes G and H of the parietal lobe (see Sub-Figures  G and H). These electrodes presented the most constant values and had a magnitude around $\approx$ 3.1 to 3.2 in awake resting-state conditions. While during general anesthesia, the Permutation Entropy was widespread over the range from $\approx$ 2.2 to 3.2, not being noticed prominent overlap among the distributions of both conditions. 
}
\label{figure2}
\end{figure}
\begin{multicols}{2}


\end{multicols}
\begin{figure*}[!h]
  \includegraphics[width=\textwidth]
  {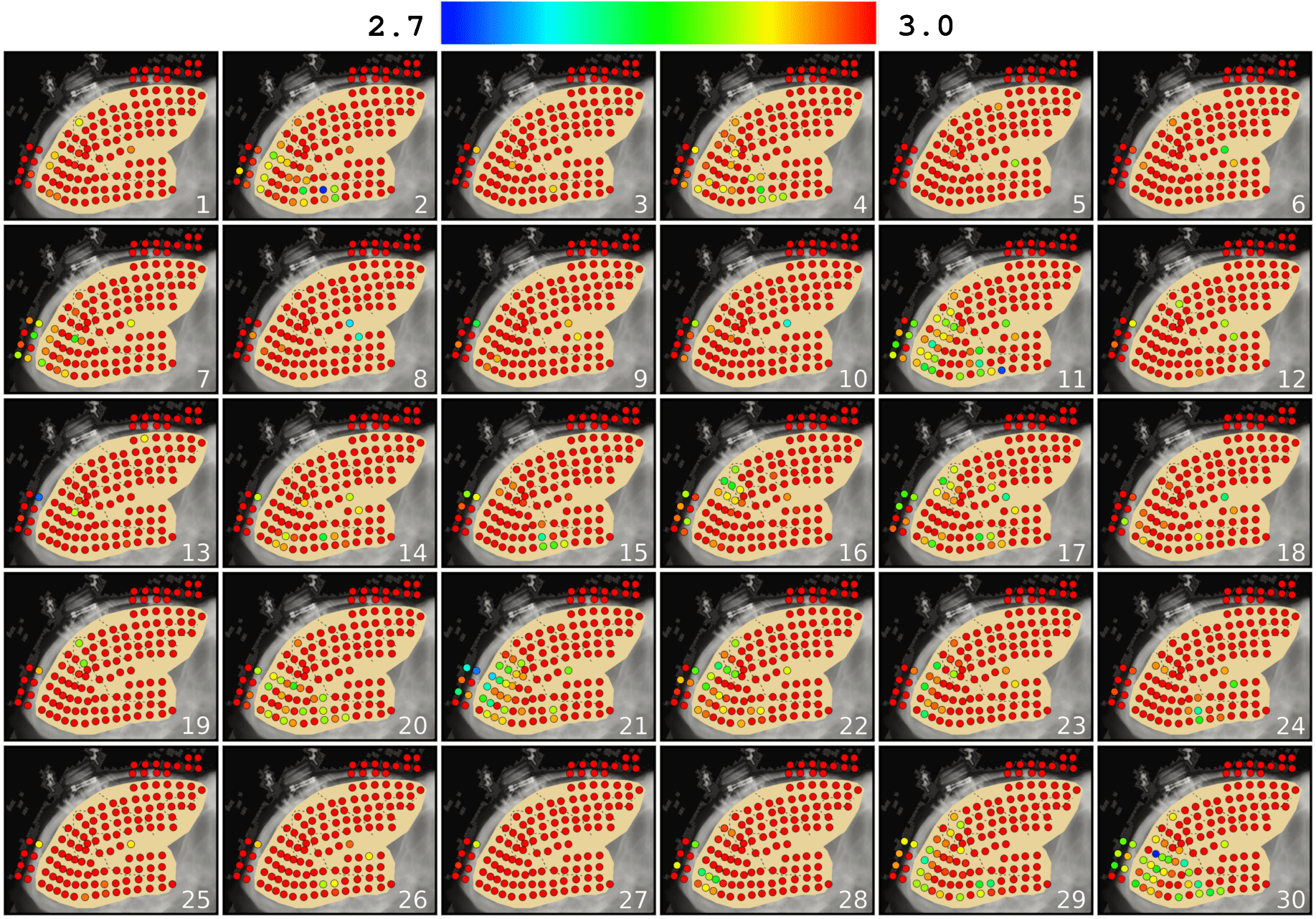}
     \caption{\textbf{Permutation Entropy during awake resting-state conditions}. In this Figure, it is shown the permutation Entropy values over the locations of their respective electrodes. Sub-Figures were estimated sequentially throughout time at every 2.0 seconds, being this the time interval between each Sub-Figure and its subsequent one. The color gradient indicates the magnitude of the Permutation Entropy. In this Figure, it is possible to visualize the characteristic pattern of the complexity in awakened resting-state conditions. It is noticeable that under these conditions, most of the electrodes present a Permutation Entropy value of around 3.0 and over, and the whole cortex did not display an expressive variation over time, as the Sub-Figures display one main characteristic pattern. Nonetheless, it is possible to observe that electrodes located in the occipital regions showed some tendency to variation (see $Sub$-$Figures$ 2, 20, 21, and 30). Considering the cortex as a whole, the region less prone to variation was the frontal lobe.
}
   \hypertarget{FIGURE3}{}  
 \end{figure*}
 \begin{multicols}{2}


\end{multicols}
\begin{figure*}[!h]
  \includegraphics[width=\textwidth]
  {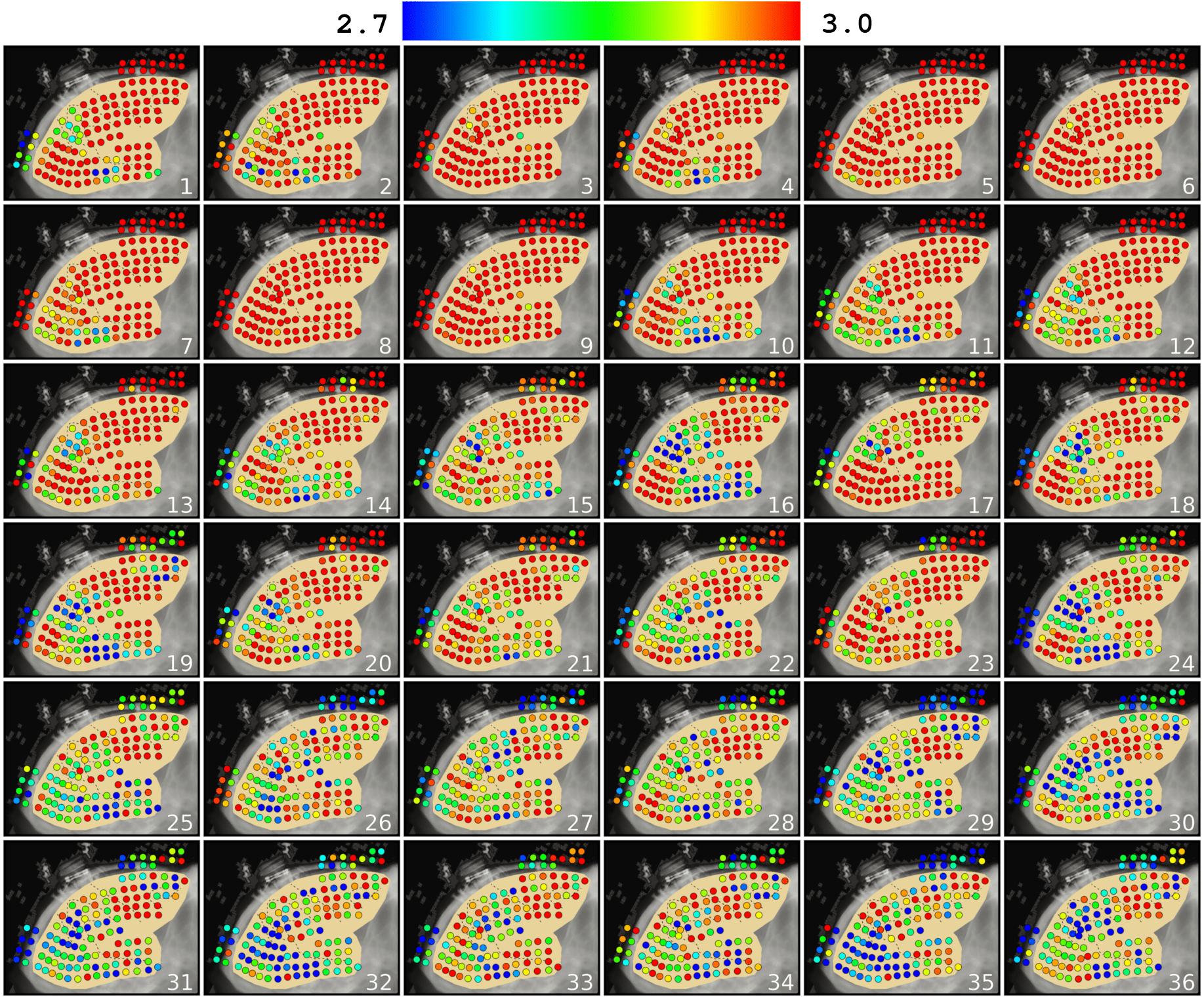}
     \caption{\textbf{Permutation Entropy during the transition to Ketamine-Medetomidine induced unconsciousness.} 
This Figure shows the Permutation Entropy values over the locations of their respective electrodes. Sub-Figures were estimated sequentially throughout time at every 2.0 seconds, being this the time interval between each Sub-Figure and its subsequent one. The color gradient indicates the magnitude of the Permutation Entropy.
In this Figure, it is possible to visualize the first alterations that happened in the patterns of the awake resting-state following the administration of the anesthetics. Primarily some regions of the temporal and occipital lobes presented a reduction, while frontal electrodes did not display expressive alterations (see $Sub$-$Figures$ 10, 14, 15, 16, and 19). Thereafter, from Sub-Figure 26 onwards, a reduction in the electrodes of the frontal lobe was verified. A remarkable characteristic observed during these first moments of the transition was that the central sulcus and nearby areas tended to exhibit high Permutation Entropy (see $Sub$-$Figures$ 15 to 36).
$Sub$-$Figures$ 8 and 9 had patterns characteristic of awake conditions, whereas $Sub$-$Figures$ 28 and 29 already displayed substantially different features. Considering that the time interval among consecutive Sub-Figures is 2.0 seconds, we infer that the first abrupt changes in the complexity of the electrophysiological activity of the cortex as a whole took place within about 30 to 40 seconds.
}
     \hypertarget{FIGURE4}{}
 \end{figure*}
 \begin{multicols}{2}


\end{multicols}
\begin{figure*}[!h]
  \includegraphics[width=\textwidth]
  {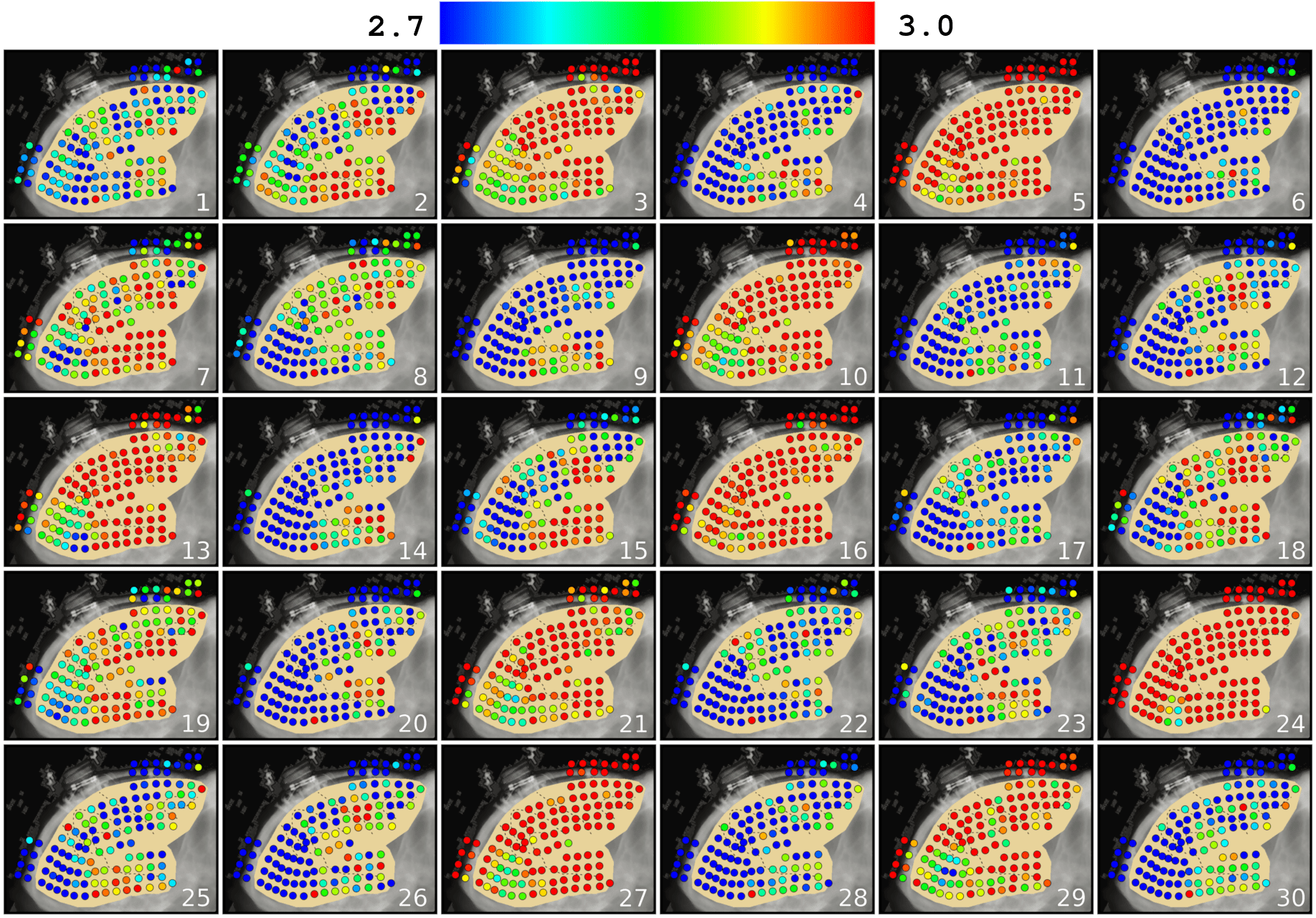}
     \caption{\textbf{Permutation Entropy during general anesthesia.} This Figure shows the Permutation Entropy values over the locations of their respective electrodes. Sub-Figures were estimated sequentially throughout time at every 2.0 seconds, being this the time interval between each Sub-Figure and its subsequent one. The color gradient indicates the magnitude of the Permutation Entropy.
It is possible to visualize the characteristic patterns and dynamics of Permutation Entropy during the state of general anesthesia. It is noticeable that the cortex alternates between high complexity states that resemble those observed during alert conditions (see $Sub$-$Figures$ $3,\: 5,\: 10,\: 16,\: 24,\: 27$ and $29$), and low complexity states in which the vast majority of cortical areas display reduced Permutation Entropy values (see $Sub$-$Figures$ 4,  6, 9, 11, 12, 14, 17, 20, 22, 23, 25, 26, 28 and 30). It was also observed that even at periods of low complexity, there was still a tendency for the temporal lobe and the central sulcus to assume slightly higher values than the rest of the cortex (see $Sub$-$Figures$ 2, 7, 12, 15, 18, 19, and 26), being a recurrent pattern that appeared with a relative frequency over general anesthesia conditions, although not always present throughout all the time.
 }
     \hypertarget{FIGURE5}{}
 \end{figure*}
 \begin{multicols}{2}


\end{multicols}
\begin{figure}[!h]
\begin{subfigure}{.32\textwidth}
  \centering
  \includegraphics[width=1\linewidth]{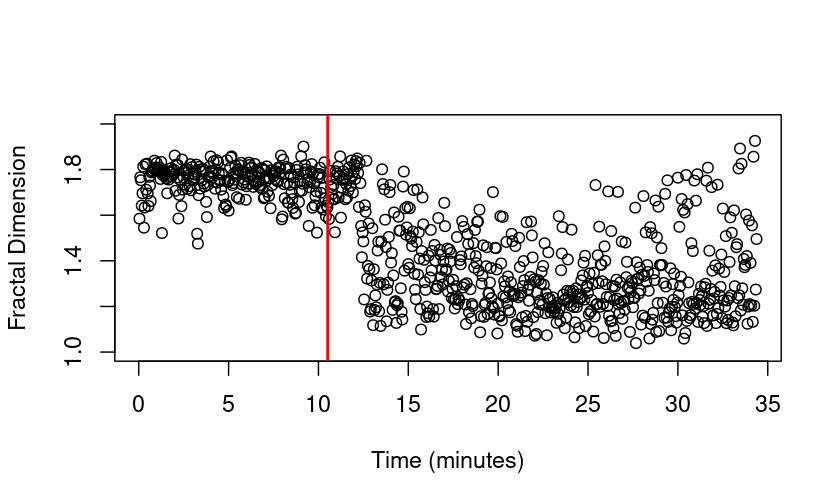}
  \caption{Electrode A}
  \label{fig6:sfigA}
\end{subfigure}%
\begin{subfigure}{.32\textwidth}
  \centering
  \includegraphics[width=1\linewidth]{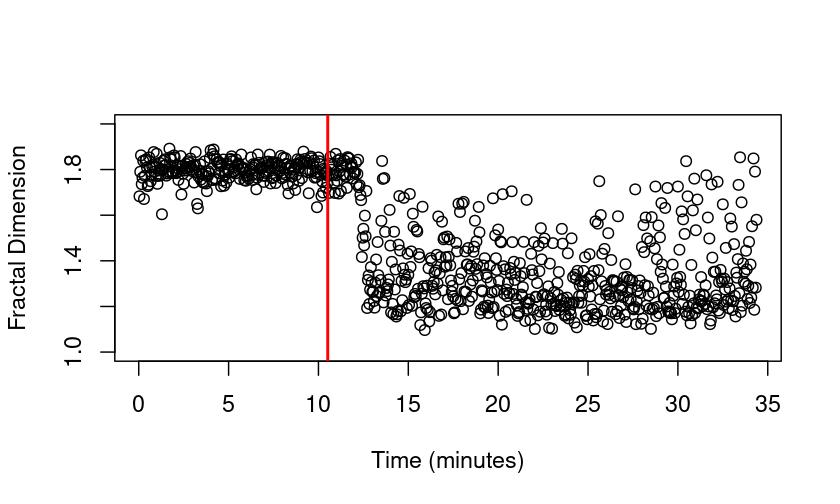}
  \caption{Electrode B}
  \label{fig6:sfigB}
\end{subfigure}%
\begin{subfigure}{.32\textwidth}
  \centering
  \includegraphics[width=1\linewidth]{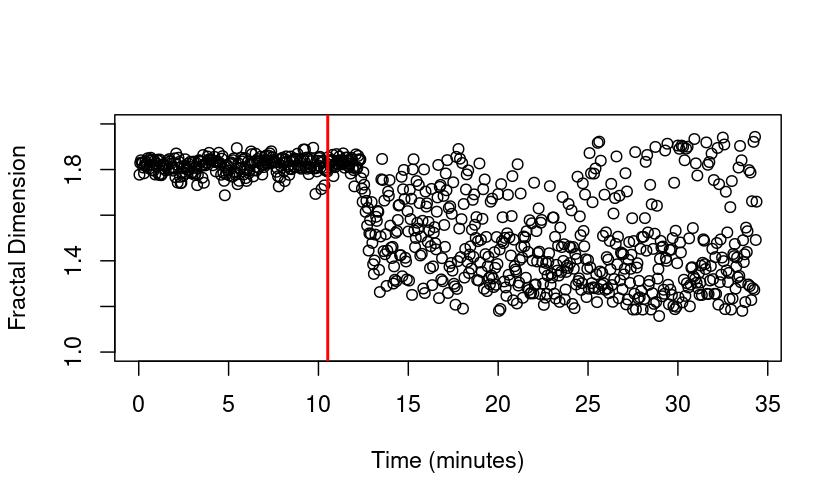}
  \caption{Electrode C}
  \label{fig6:sfigC}
\end{subfigure}\\%
\begin{subfigure}{.32\textwidth}
  \centering
  \includegraphics[width=1\linewidth]{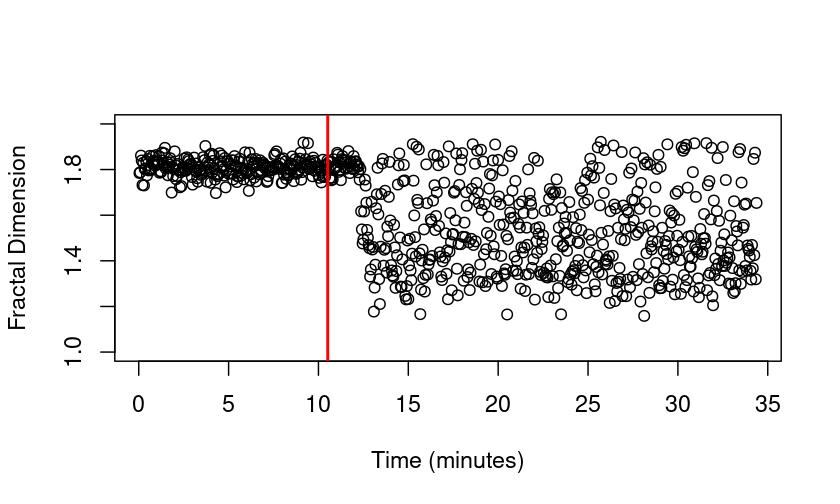}
  \caption{Electrode D}
  \label{fig6:sfigD}
\end{subfigure}%
\begin{subfigure}{.32\textwidth}
  \centering
  \includegraphics[width=1\linewidth]{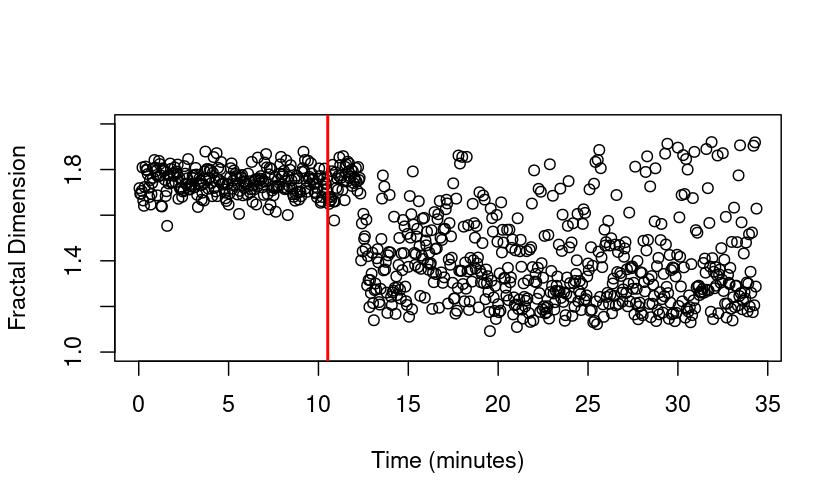}
  \caption{Electrode E}
  \label{fig6:sfigE}
\end{subfigure}%
\begin{subfigure}{.32\textwidth}
  \centering
  \includegraphics[width=1\linewidth]{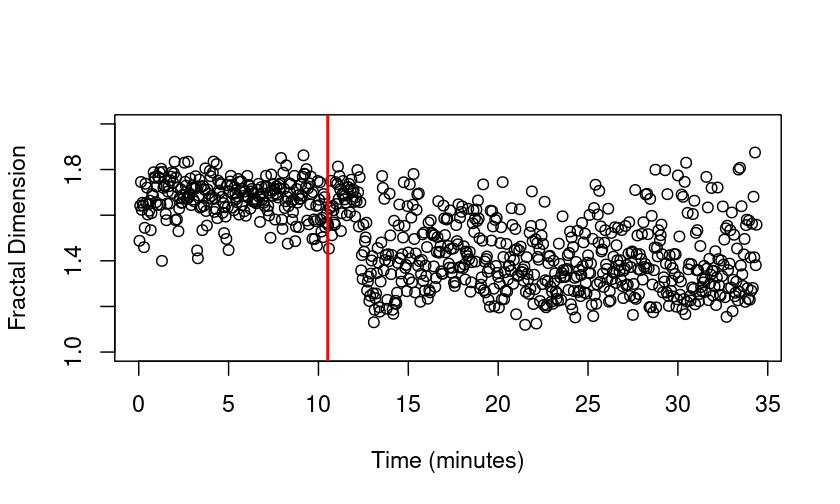}
  \caption{Electrode F}
  \label{fig6:sfigF}
\end{subfigure}\\%
\begin{subfigure}{.32\textwidth}
  \centering
  \includegraphics[width=1\linewidth]{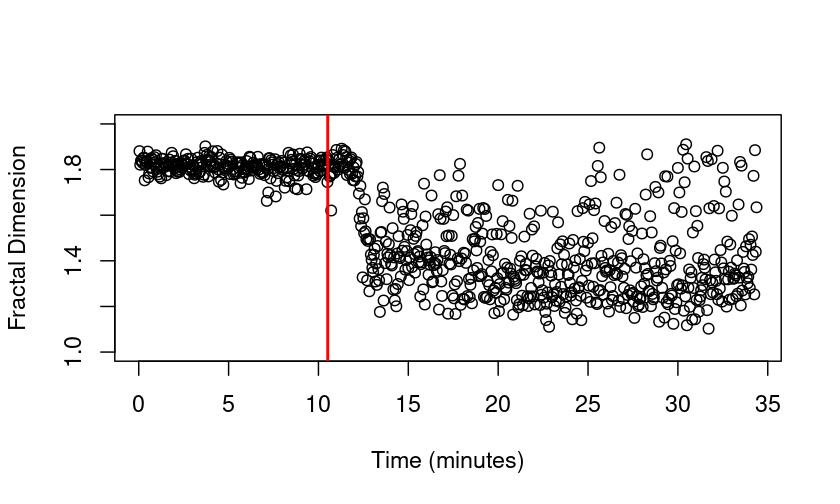}
  \caption{Electrode G}
  \label{fig6:sfigG}
\end{subfigure}%
\begin{subfigure}{.32\textwidth}
  \centering
  \includegraphics[width=1\linewidth]{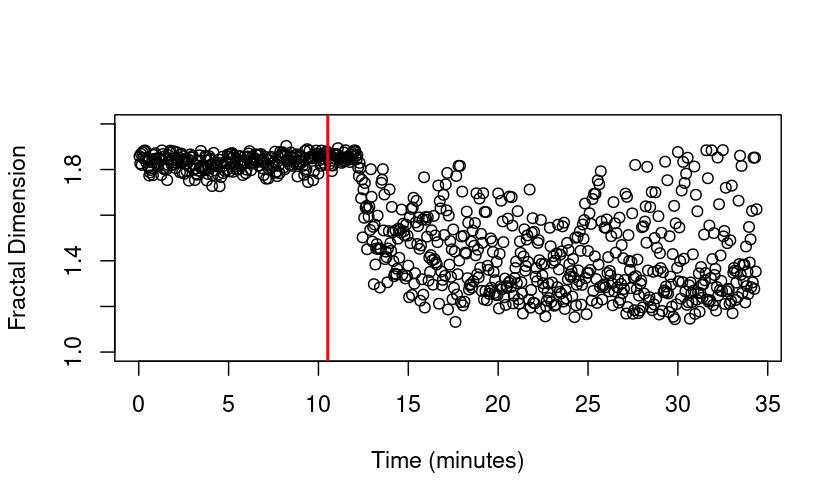}
  \caption{Electrode H}
  \label{fig6:sfigH}
\end{subfigure}%
\begin{subfigure}{.32\textwidth}
  \centering
  \includegraphics[width=1\linewidth]{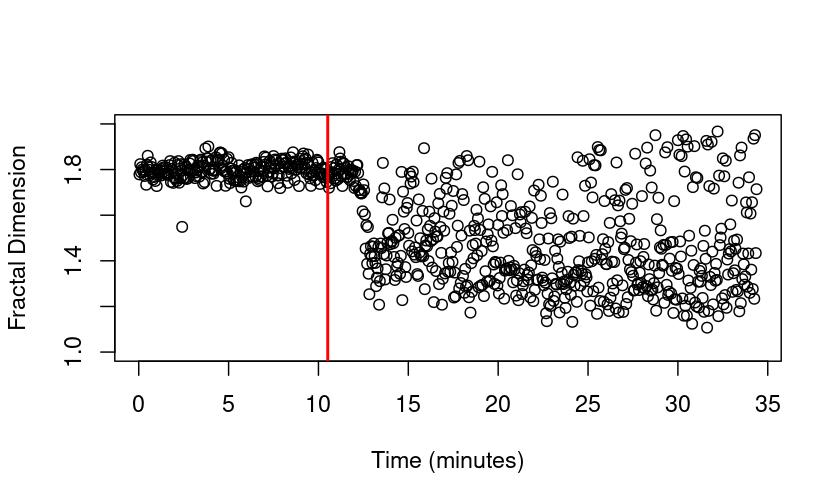}
  \caption{Electrode I}
  \label{fig6:sfigI}
\end{subfigure}\\%
\begin{subfigure}{.32\textwidth}
  \centering
  \includegraphics[width=1\linewidth]{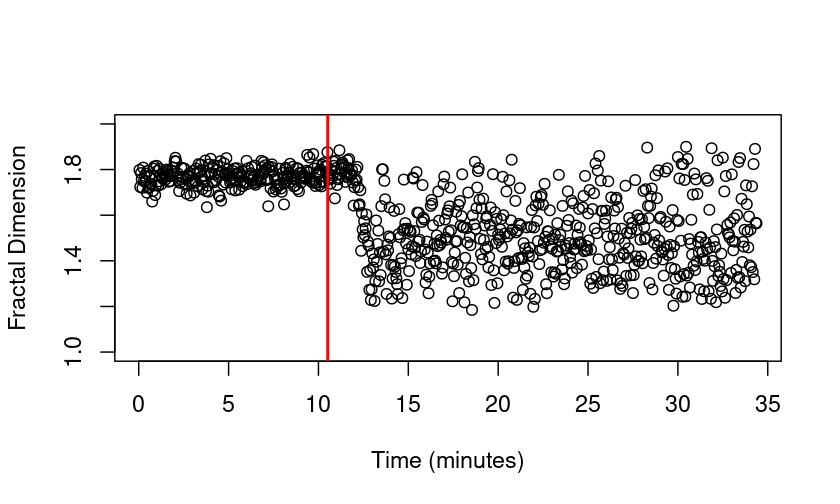}
  \caption{Electrode J}
  \label{fig6:sfigJ}
\end{subfigure}%
\begin{subfigure}{.32\textwidth}
  \centering
  \includegraphics[width=1\linewidth]{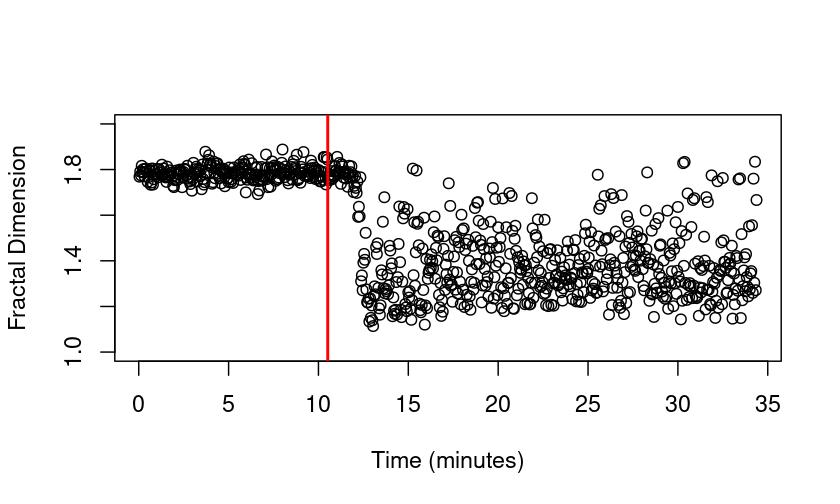}
  \caption{Electrode K}
  \label{fig6:sfigK}
\end{subfigure}%
\begin{subfigure}{.32\textwidth}
  \centering
  \includegraphics[width=1\linewidth]{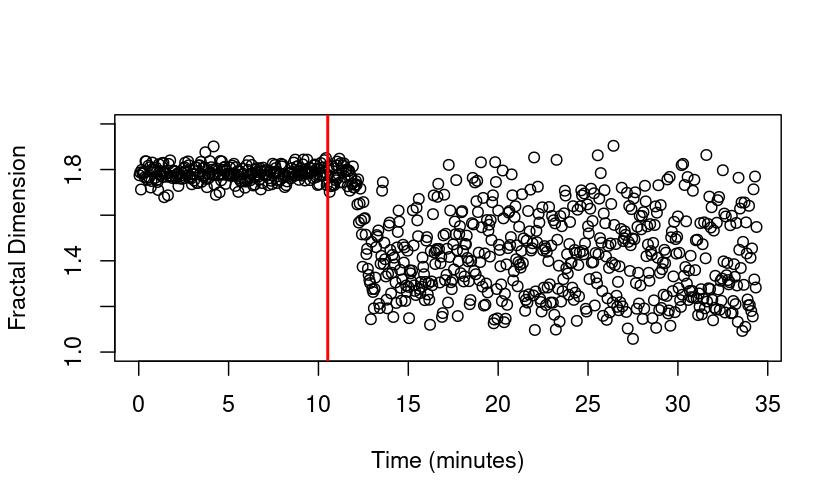}
  \caption{Electrode L}
  \label{fig6:sfigL}
\end{subfigure}\\%
\begin{subfigure}{.32\textwidth}
  \centering
  \includegraphics[width=1\linewidth]{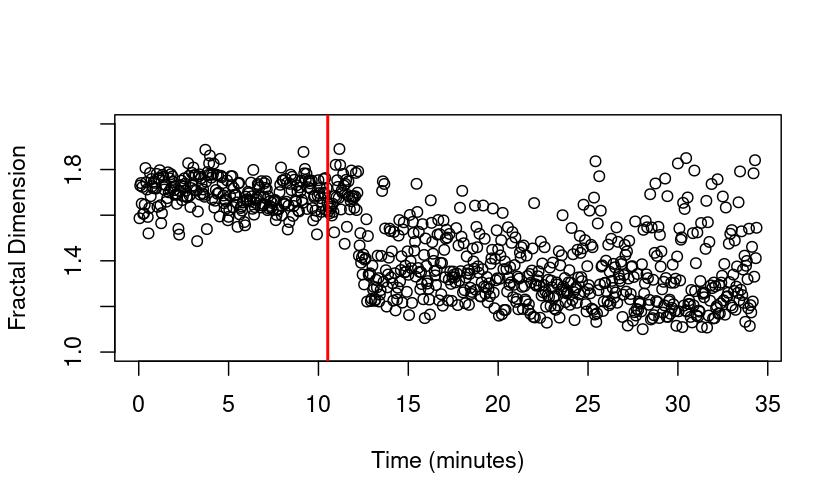}
  \caption{Electrode M}
  \label{fig6:sfigM}
\end{subfigure}%
\begin{subfigure}{.32\textwidth}
  \centering
  \includegraphics[width=1\linewidth]{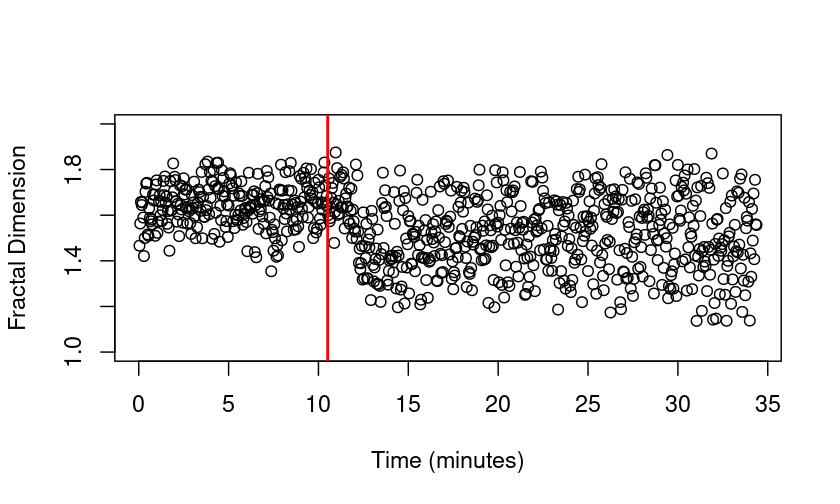}
  \caption{Electrode N}
  \label{fig6:sfigN}
\end{subfigure}%
\begin{subfigure}{.32\textwidth}
  \centering
  \includegraphics[width=1\linewidth]{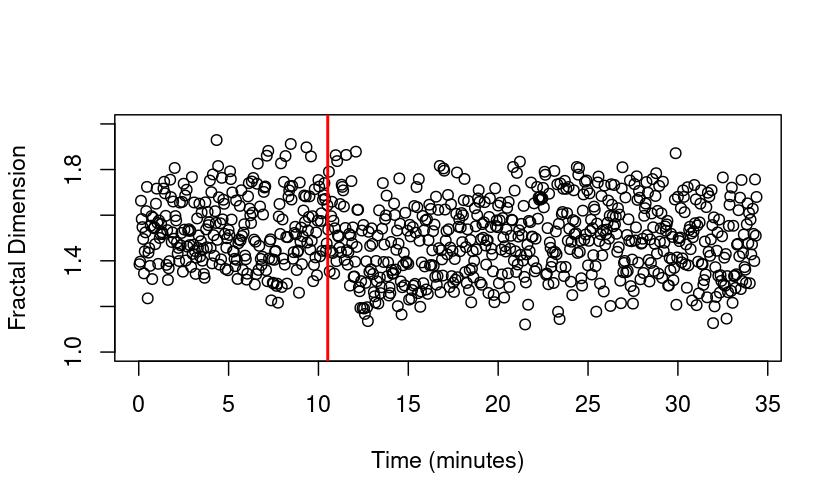}
  \caption{Electrode O}
  \label{fig6:sfigO}
\end{subfigure}\\%
\caption{\textbf{Fractal Dimension of the electrophysiological cortical activity throughout the anesthetic induction.}}
\label{figure6A}
\hypertarget{FIGURE6}{}
\end{figure}
\begin{multicols}{2}


\end{multicols}
\begin{figure}[!h]\ContinuedFloat
\begin{subfigure}{.32\textwidth}
  \centering
  \includegraphics[width=1\linewidth]{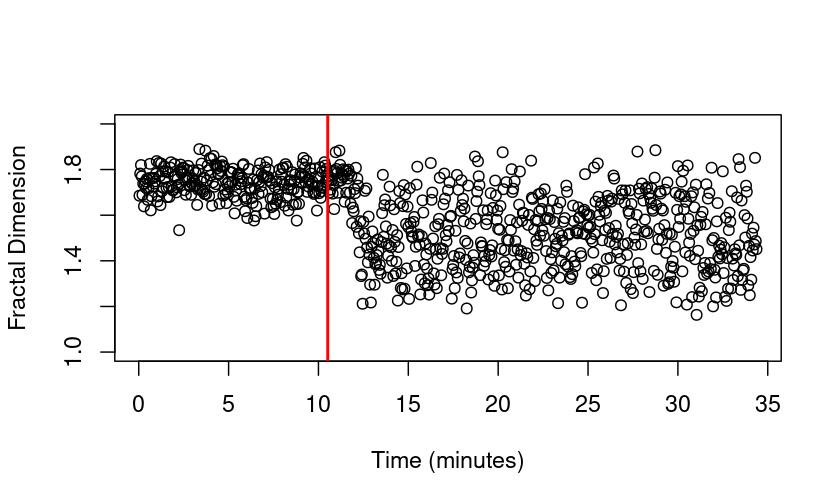}
  \caption{Electrode P}
  \label{fig6:sfigP}
\end{subfigure}%
\begin{subfigure}{.32\textwidth}
  \centering
  \includegraphics[width=1\linewidth]{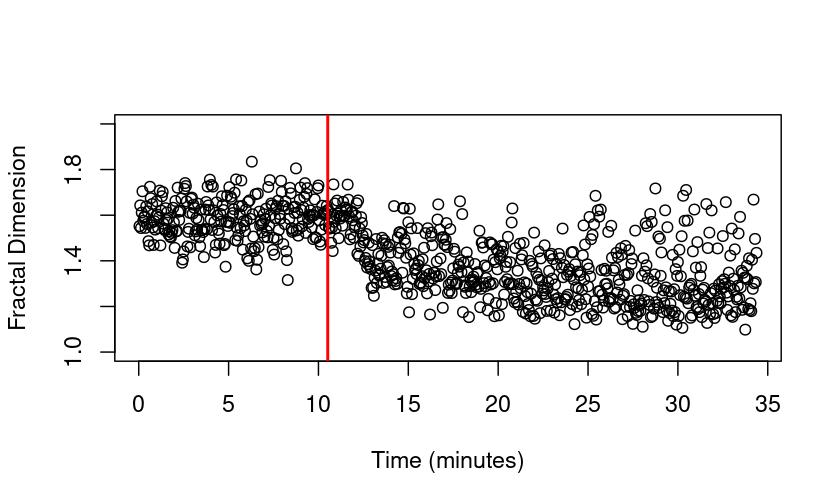}
  \caption{Electrode Q}
  \label{fig6:sfigQ}
\end{subfigure}%
\begin{subfigure}{.32\textwidth}
  \centering
  \includegraphics[width=1\linewidth]{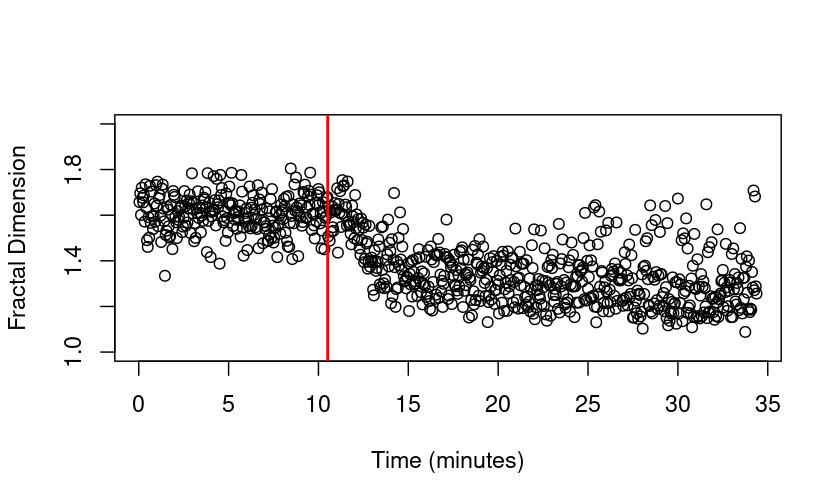}
  \caption{Electrode R}
  \label{fig6:sfigR}
\end{subfigure}\\%
\begin{subfigure}{.32\textwidth}
  \centering
  \includegraphics[width=1\linewidth]{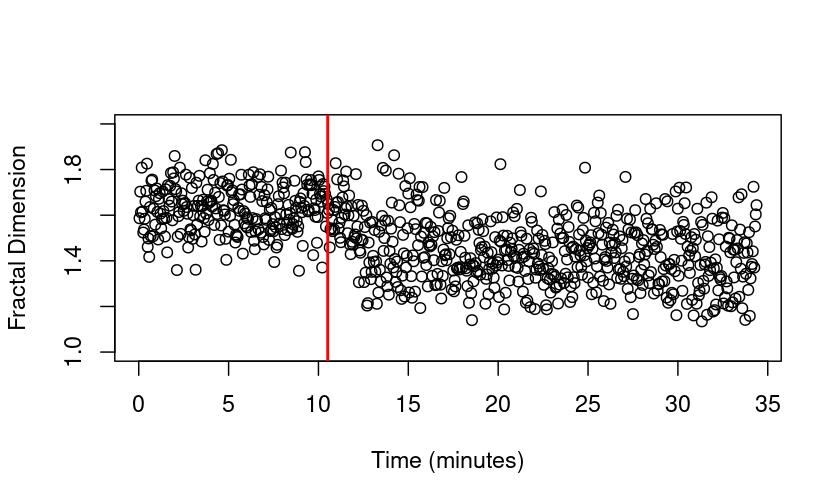}
  \caption{Electrode S}
  \label{fig6:sfigS}
\end{subfigure}%
\begin{subfigure}{.32\textwidth}
  \centering
  \includegraphics[width=1\linewidth]{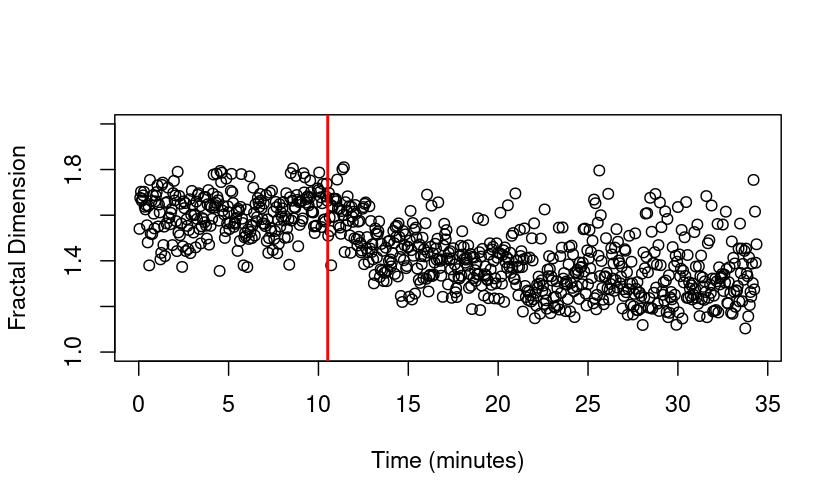}
  \caption{Electrode T}
  \label{fig6:sfigT}
\end{subfigure}%
\begin{subfigure}{.32\textwidth}
  \centering
  \includegraphics[width=1\linewidth]{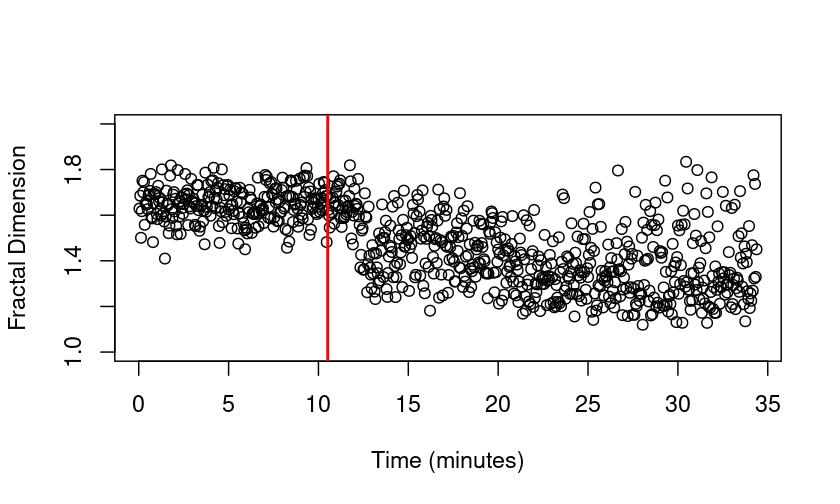}
  \caption{Electrode U}
  \label{fig6:sfigU}
\end{subfigure}\\%
\begin{subfigure}{.32\textwidth}
  \centering
  \includegraphics[width=1\linewidth]{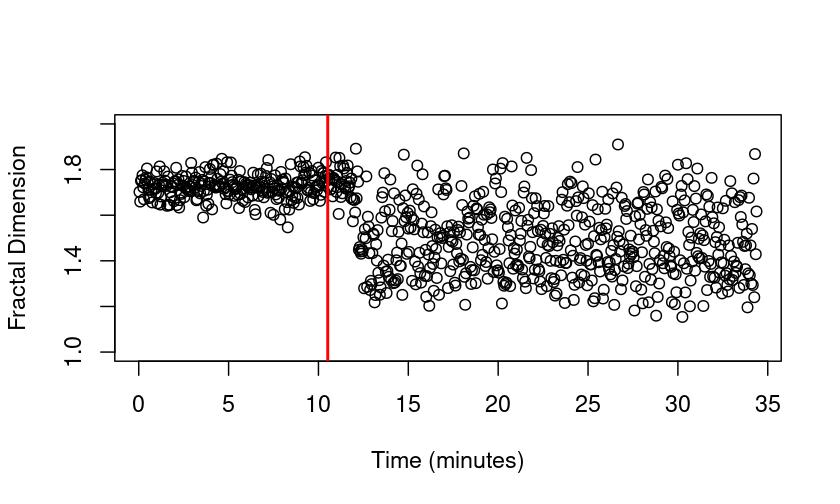}
  \caption{Electrode V}
  \label{fig6:sfigV}
\end{subfigure}

\caption{\textbf{Fractal Dimension of the electrophysiological cortical activity throughout the anesthetic induction.}
Each Sub-Figure is respective to an electrode that was positioned over a specific cortical area, for the exact location of each electrode (see \protect\hyperlink{FIGURE12}{Supplementary $Figure \cdot 1$}, the corresponding letters from A to V). Brain activity started to be recorded in awakening resting-state conditions. The anesthetic drugs cocktail was administrated at 10.5 minutes; this event is indicated in each Sub-Figure by a vertical red line. The Fractal Dimension was calculated every 2.5 seconds throughout the experiment, being this the time interval between each point and its subsequent one.
In this Figure, it is possible to observe the dynamics of the Fractal Dimension throughout the anesthetic induction experiment. During awakened alert conditions, there was a tendency to remain approximately constant around  $\approx$ 1.7 to 1.8 in most electrodes. However, within about 1.5 to 2.0 minutes after the anesthetic agents administration, the Fractal Dimension values presented an abrupt change; they decreased and started to assume a considerably more significant variation than in alert conditions, apparently fluctuating randomly between a minimum of approximately 1.3 and a maximum of about 1.8. Furthermore, it was noticeable that the behavior of the Fractal Dimension was not the same over different cortical regions, evidencing that the effects of the anesthetics on the electrophysiological activity are not the same throughout the whole cortex.
}
\label{figure6B}
\end{figure}
\begin{multicols}{2}


\end{multicols}
\begin{figure}[h]
\begin{subfigure}{.245\textwidth}
  \centering
  \includegraphics[width=1\linewidth]{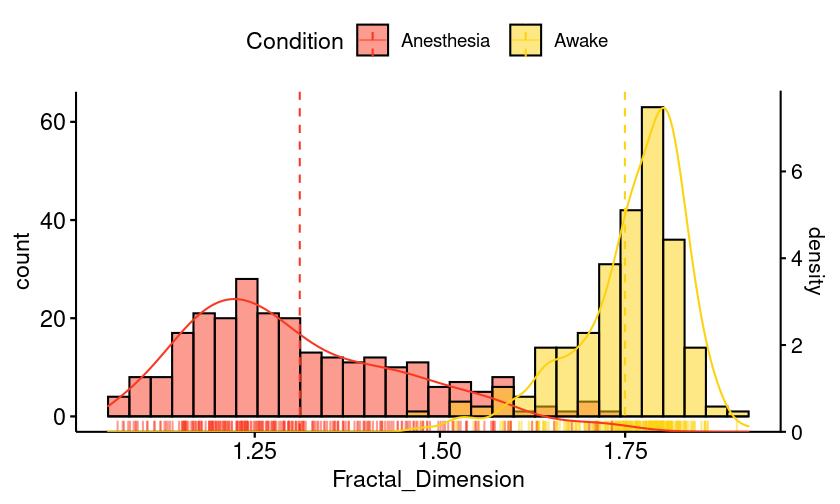}
  \caption{Electrode A}
  \label{fig7:sfigA}
\end{subfigure}%
\begin{subfigure}{.245\textwidth}
  \centering
  \includegraphics[width=1\linewidth]{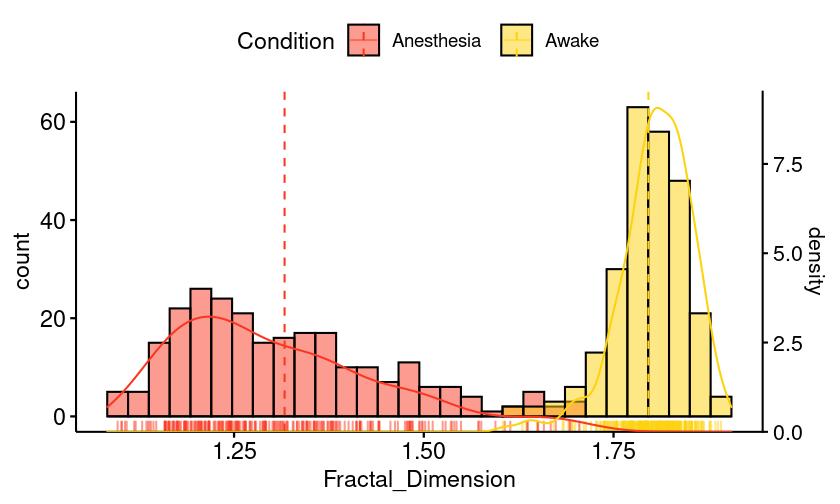}
  \caption{Electrode B}
  \label{fig7:sfigB}
\end{subfigure}%
\begin{subfigure}{.245\textwidth}
  \centering
  \includegraphics[width=1\linewidth]{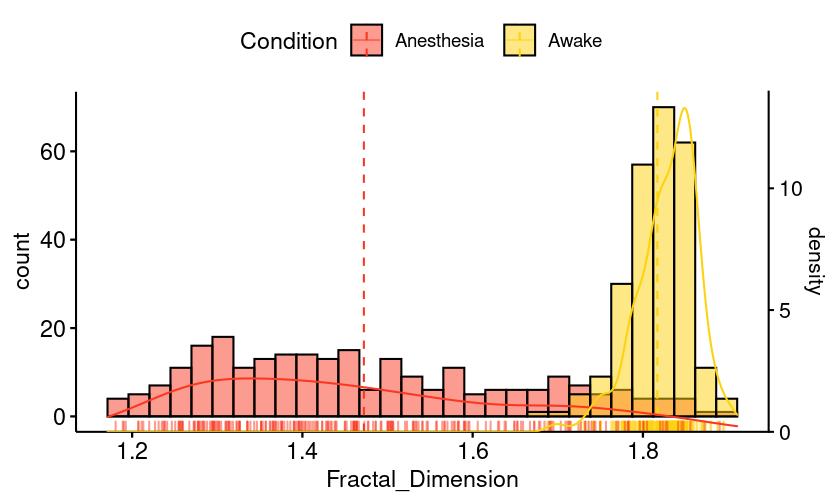}
  \caption{Electrode C}
  \label{fig7:sfigC}
\end{subfigure}%
\begin{subfigure}{.245\textwidth}
  \centering
  \includegraphics[width=1\linewidth]{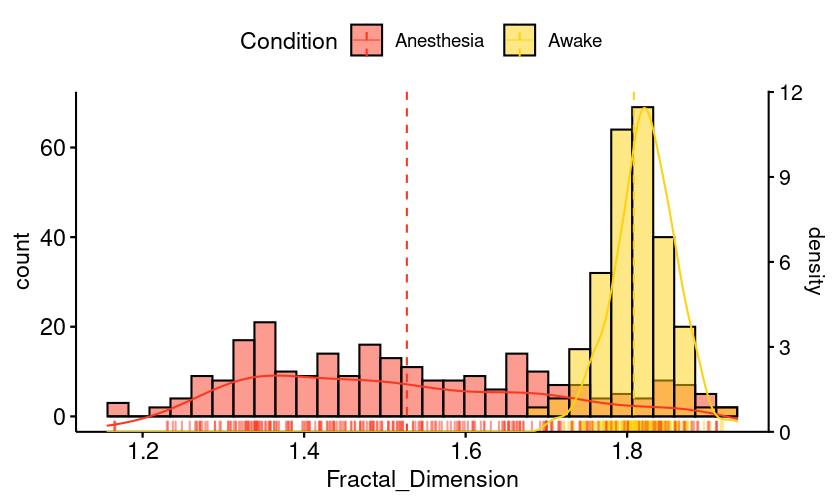}
  \caption{Electrode D}
  \label{fig7:sfigD}
\end{subfigure}\\%
\begin{subfigure}{.245\textwidth}
  \centering
  \includegraphics[width=1\linewidth]{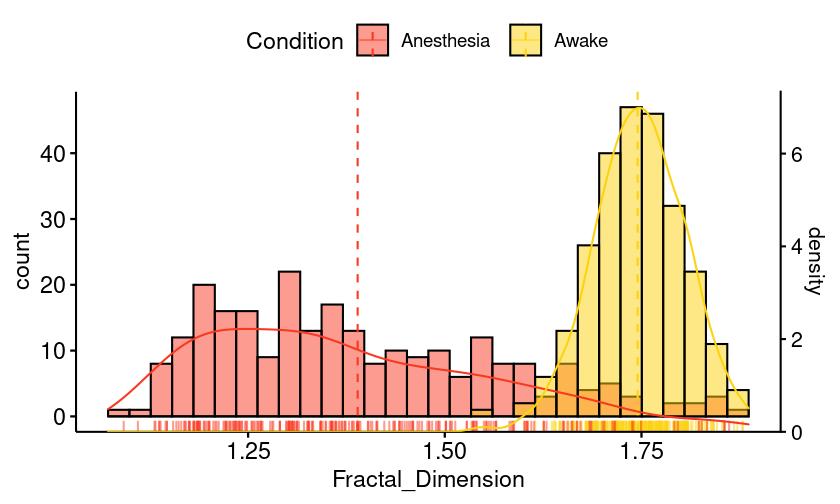}
  \caption{Electrode E}
  \label{fig7:sfigE}
\end{subfigure}%
\begin{subfigure}{.245\textwidth}
  \centering
  \includegraphics[width=1\linewidth]{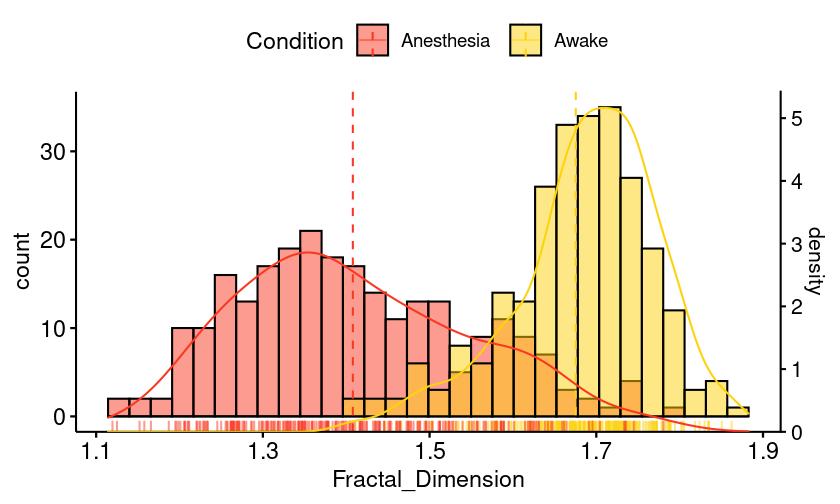}
  \caption{Electrode F}
  \label{fig7:sfigF}
\end{subfigure}%
\begin{subfigure}{.245\textwidth}
  \centering
  \includegraphics[width=1\linewidth]{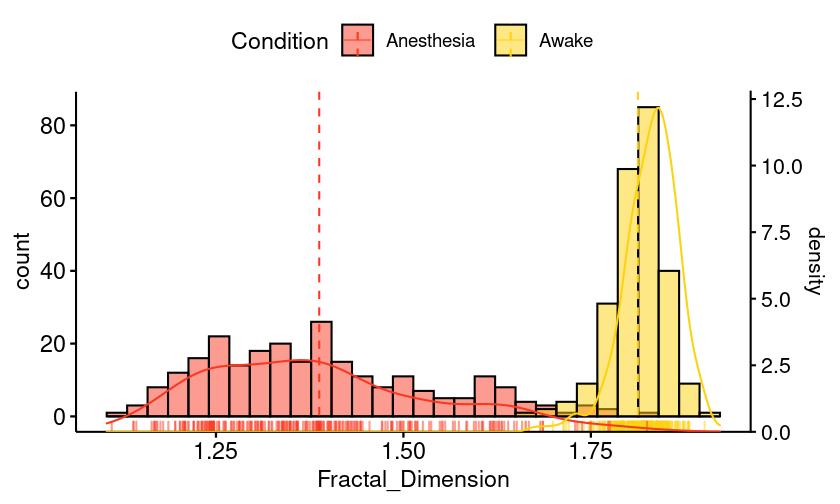}
  \caption{Electrode G}
  \label{fig7:sfigG}
\end{subfigure}%
\begin{subfigure}{.245\textwidth}
  \centering
  \includegraphics[width=1\linewidth]{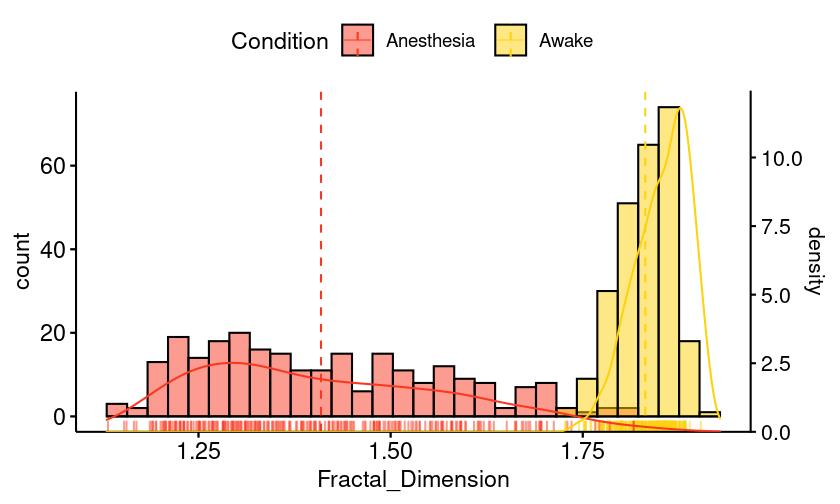}
  \caption{Electrode H}
  \label{fig7:sfigH}
\end{subfigure}\\%
\begin{subfigure}{.245\textwidth}
  \centering
  \includegraphics[width=1\linewidth]{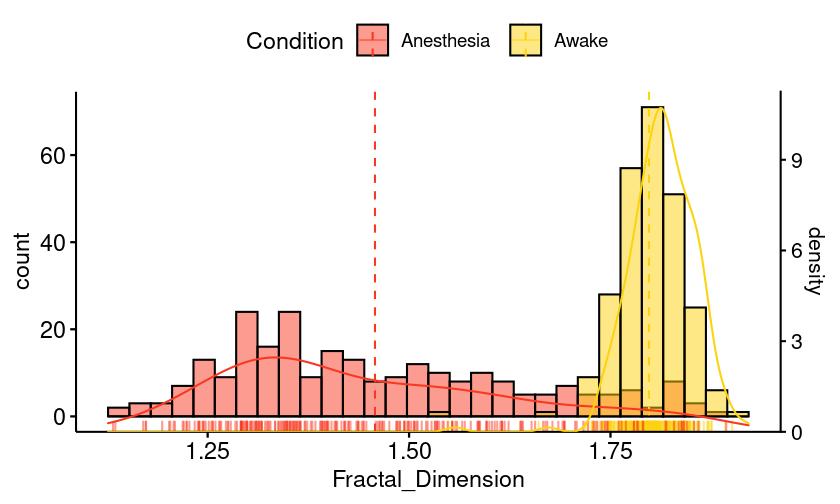}
  \caption{Electrode I}
  \label{fig7:sfigI}
\end{subfigure}%
\begin{subfigure}{.245\textwidth}
  \centering
  \includegraphics[width=1\linewidth]{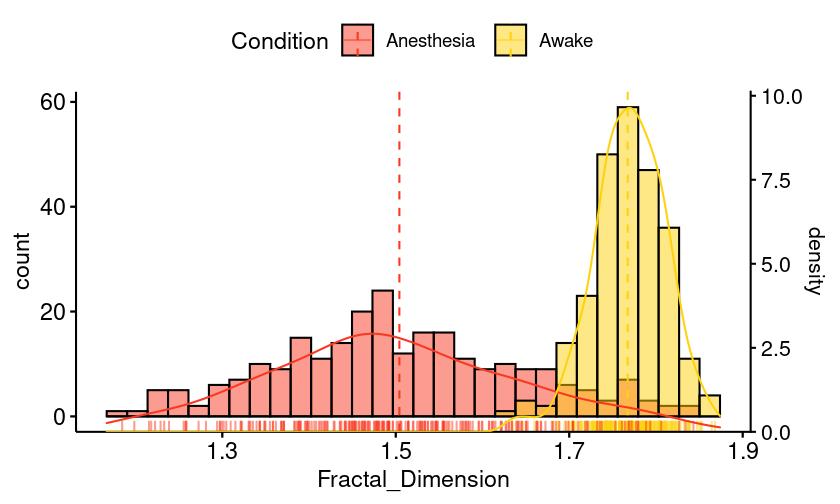}
  \caption{Electrode J}
  \label{fig7:sfigJ}
\end{subfigure}%
\begin{subfigure}{.245\textwidth}
  \centering
  \includegraphics[width=1\linewidth]{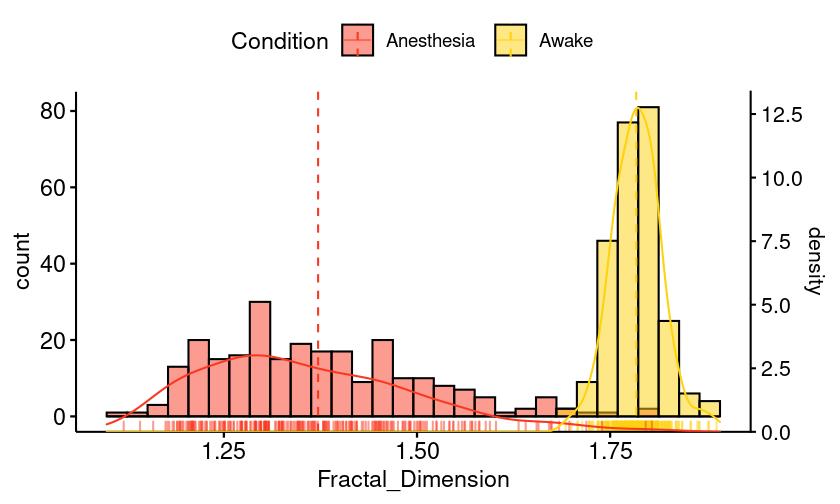}
  \caption{Electrode K}
  \label{fig7:sfigK}
\end{subfigure}%
\begin{subfigure}{.245\textwidth}
  \centering
  \includegraphics[width=1\linewidth]{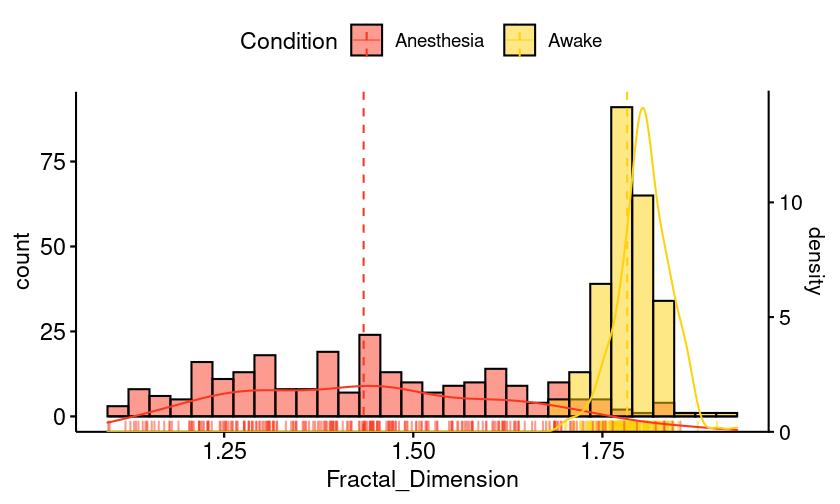}
  \caption{Electrode L}
  \label{fig7:sfigL}
\end{subfigure}\\%
\begin{subfigure}{.245\textwidth}
  \centering
  \includegraphics[width=1\linewidth]{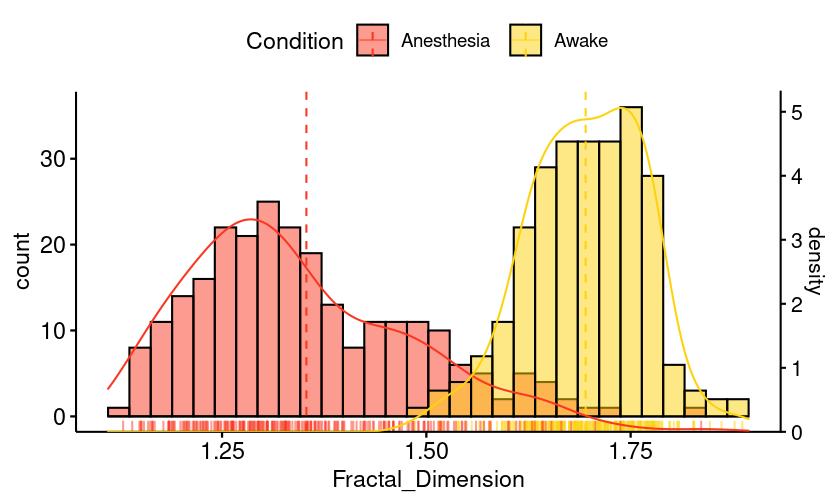}
  \caption{Electrode M}
  \label{fig7:sfigM}
\end{subfigure}%
\begin{subfigure}{.245\textwidth}
  \centering
  \includegraphics[width=1\linewidth]{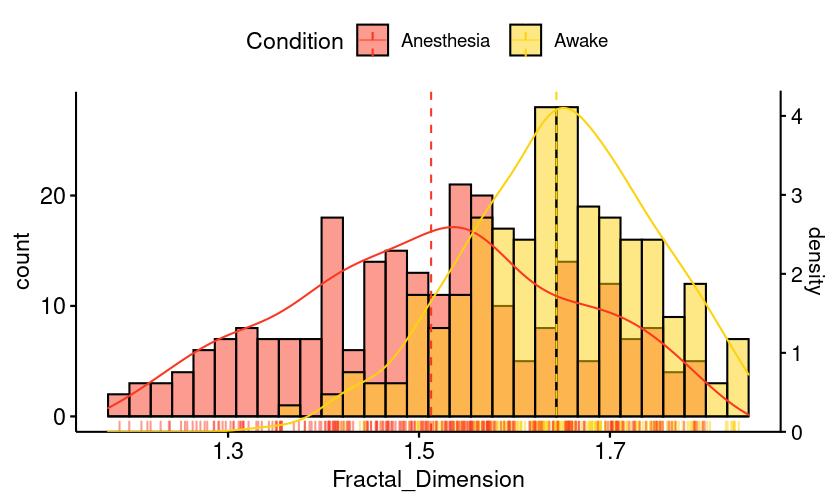}
  \caption{Electrode N}
  \label{fig7:sfigN}
\end{subfigure}%
\begin{subfigure}{.245\textwidth}
  \centering
  \includegraphics[width=1\linewidth]{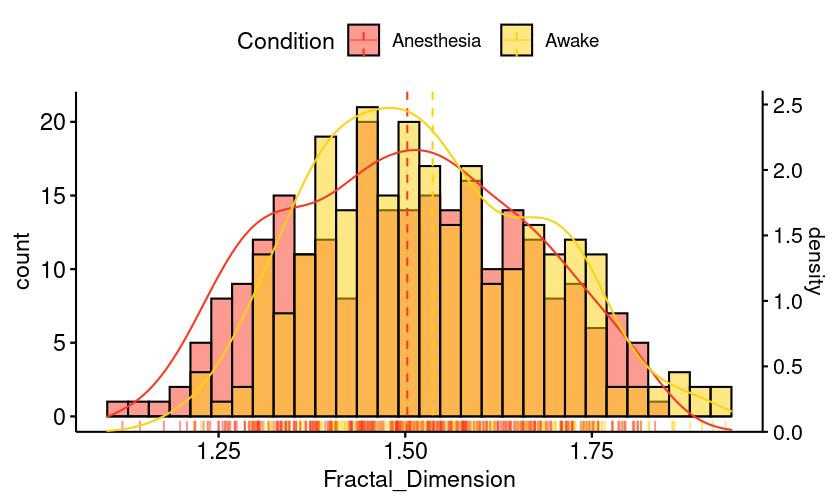}
  \caption{Electrode O}
  \label{fig7:sfigO}
\end{subfigure}%
\begin{subfigure}{.245\textwidth}
  \centering
  \includegraphics[width=1\linewidth]{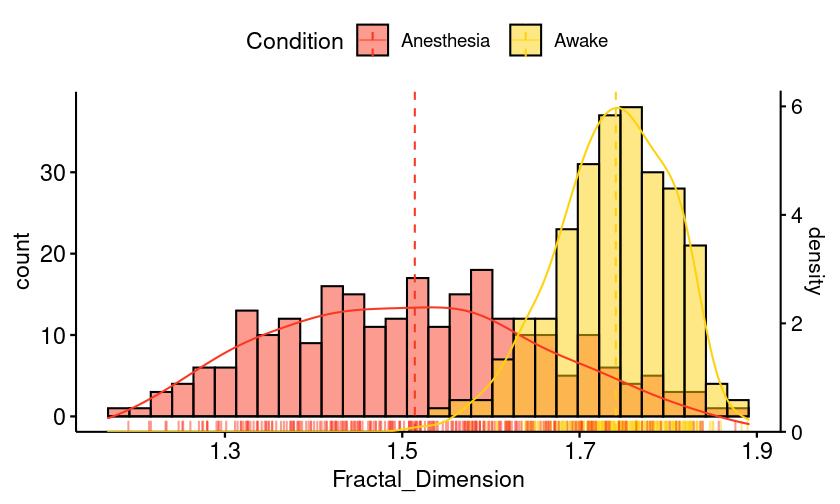}
  \caption{Electrode P}
  \label{fig7:sfigP}
\end{subfigure}\\%
\begin{subfigure}{.245\textwidth}
  \centering
  \includegraphics[width=1\linewidth]{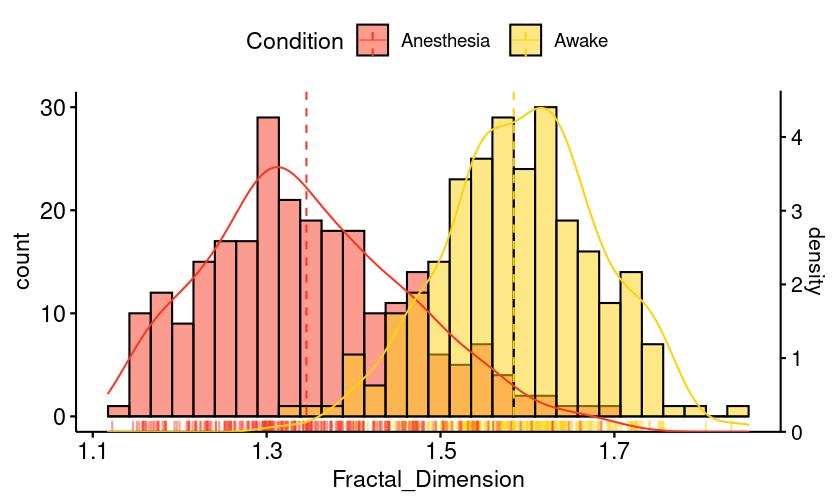}
  \caption{Electrode Q}
  \label{fig7:sfigQ}
\end{subfigure}%
\begin{subfigure}{.245\textwidth}
  \centering
  \includegraphics[width=1\linewidth]{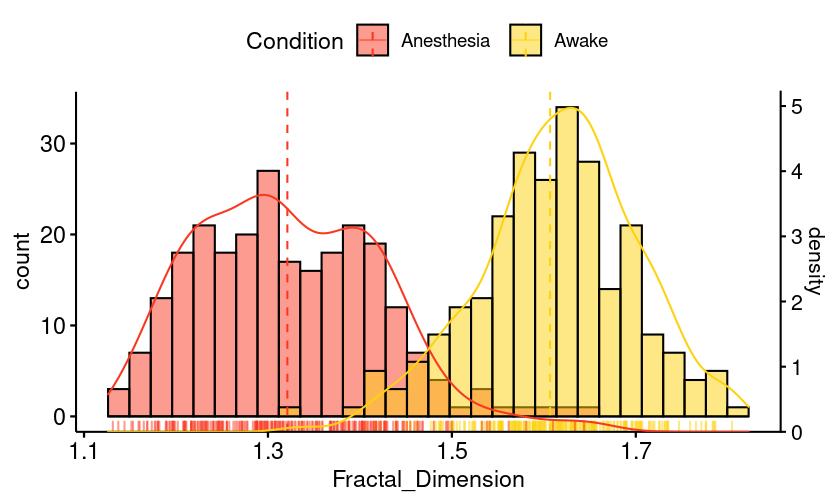}
  \caption{Electrode R}
  \label{fig7:sfigR}
\end{subfigure}%
\begin{subfigure}{.245\textwidth}
  \centering
  \includegraphics[width=1\linewidth]{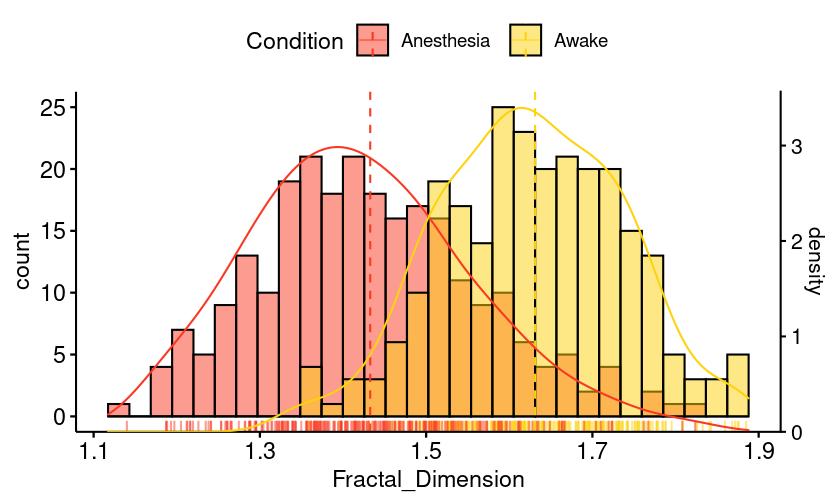}
  \caption{Electrode S}
  \label{fig7:sfigS}
\end{subfigure}%
\begin{subfigure}{.245\textwidth}
  \centering
  \includegraphics[width=1\linewidth]{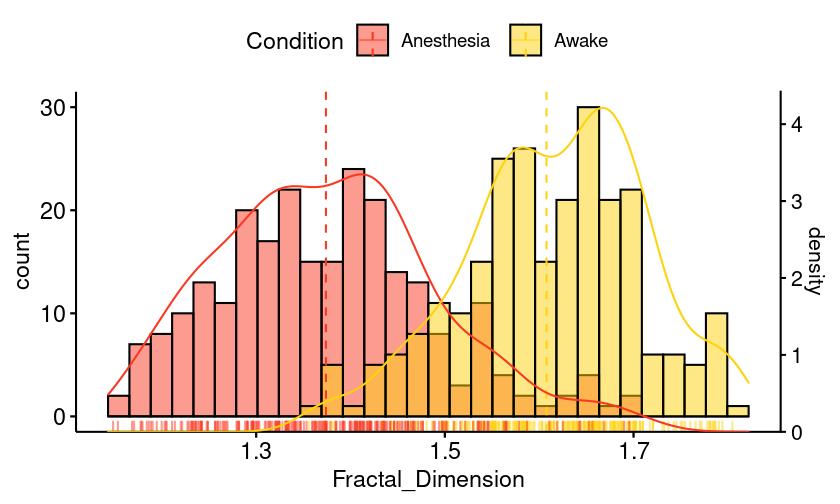}
  \caption{Electrode T}
  \label{fig7:sfigT}
\end{subfigure}\\%
\begin{subfigure}{.245\textwidth}
  \centering
  \includegraphics[width=1\linewidth]{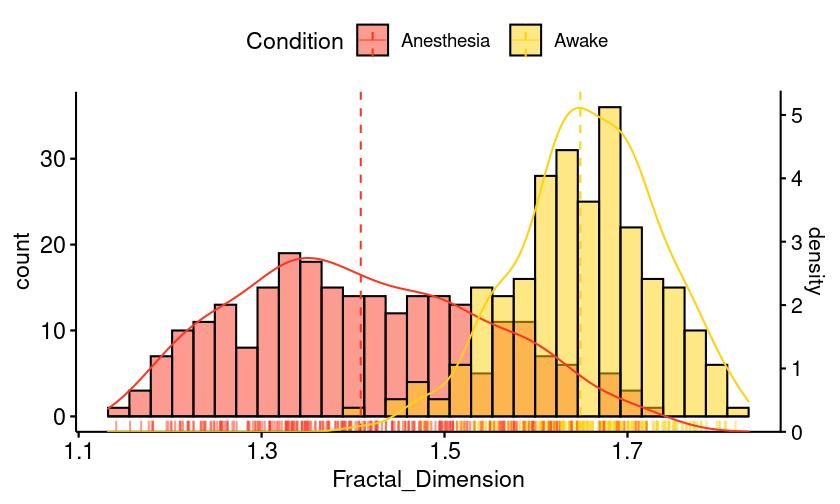}
  \caption{Electrode U}
  \label{fig7:sfigU}
\end{subfigure}%
\begin{subfigure}{.245\textwidth}
  \centering
  \includegraphics[width=1\linewidth]{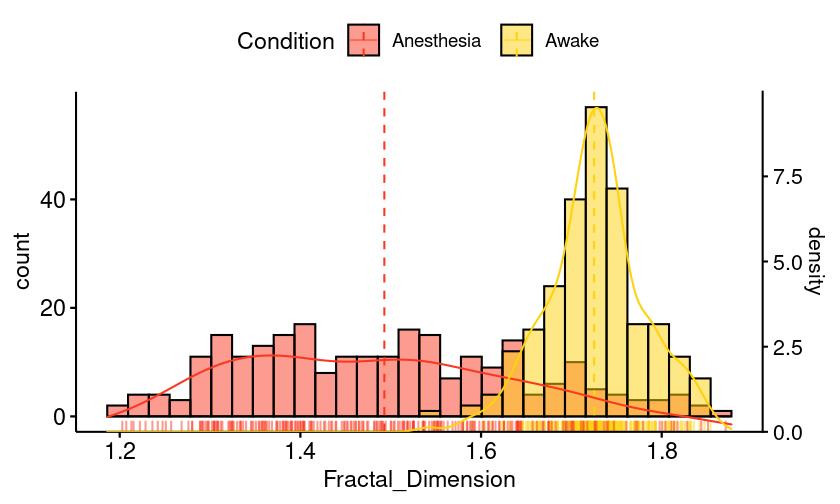}
  \caption{Electrode V}
  \label{fig7:sfigV}
\end{subfigure}%
\caption{\textbf{Overlapping histogram plots of the Fractal Dimension.} Overlapping histogram plots of the electrophysiological time series’s Fractal Dimension respective to the awakened (indicated in yellow) and general anesthesia conditions (indicated in red); see color chart. Each Sub-Figure corresponds to an electrode that was positioned over a specific cortical region; for the location of each electrode (see \protect\hyperlink{FIGURE12}{Supplementary $Figure \cdot 1$}, the corresponding letters from A to V). 
In general lines, the Fractal Dimension values of the macaque’s cortical electrophysiological activity under general anesthesia tended to be smaller and more widespread than during awaken conditions. Some distinctions were observed according to the location of the electrodes over the cortex. The ECoG electrodes positioned over the frontoparietal regions (electrodes A to L, see Sub-Figures A to L) had considerably constant values and presented a magnitude around 1.8 in awake resting-state conditions. While during general anesthesia, the Fractal Dimension was most widespread over the range from 1.1 to 1.8, existing a relatively small overlap between the distributions of the two conditions. The electrodes located in the occipital-temporal areas (electrodes M to V, see Sub-Figures M to V) presented greater variation during the awakened resting-state when compared to the frontoparietal ones. 
}
\label{figure7}
\hypertarget{FIGURE7}{}
\end{figure}
\begin{multicols}{2}



\end{multicols}
\begin{table}[!h]
\centering
\caption{\textbf{Mean and standard deviation of the Fractal Dimension values in alert and general anesthesia.} This table presents the mean and standard deviation of the Fractal Dimension values in alert and general anesthesia conditions from the electrodes located in distinct cortical regions. For the location of each electrode (see \protect\hyperlink{FIGURE12}{Supplementary $Figure \cdot 1$}, the corresponding letters from A to V). In addition, the p-values of the Wilcoxon signed-rank test are shown in the condition that the Fractal Dimension values found during anesthesia are smaller than those observed during the alert resting state. It was verified for all the electrodes that the average of the Fractal Dimension during alert conditions is higher than the average found in general anesthesia and that the standard deviation is smaller during alertness than during anesthesia. In addition, for all electrodes, the Wilcoxon test at a p-value of $5\%$ confirmed that the Fractal Dimension of the cortical records decreased during the Ketamine-Medetomidine-induced general anesthesia.
}
\vspace{0.5cm}
\begin{tabular}{l|lr|lr|c}
 \hline
\textbf{Fractal Dimension} & \multicolumn{2}{c}{Awake} \vline & \multicolumn{2}{c}{Anesthesia} \vline & \multicolumn{1}{c}{Wilcoxon Test}\\
\hline
Electrode: & Mean & SD  & Mean & SD & P-Value [Anesthesia $<$ Awake]\\ 
\hline 
\hline 
\rowcolor{gray!20}Electrode  A & 1.75 & 0.071 & 1.32 & 0.17 & 5.3e-97  \\
 Electrode  B & 1.80 & 0.047 & 1.32 & 0.156 & 2.8e-104  \\
\rowcolor{gray!20} Electrode  C & 1.82 & 0.035 & 1.47 & 0.2 & 2.4e-76  \\
Electrode  D & 1.81 & 0.038 & 1.52 & 0.191 & 1.4e-66  \\
\rowcolor{gray!20}Electrode  E & 1.74 & 0.055 & 1.38 & 0.193 & 3.8e-80  \\
Electrode  F & 1.68 & 0.085 & 1.41 & 0.148 & 3.5e-79  \\
\rowcolor{gray!20}Electrode  G & 1.81 & 0.036 & 1.39 & 0.173 & 5.6e-96  \\
Electrode  H & 1.83 & 0.034 & 1.41 & 0.177 & 8.6e-99  \\
\rowcolor{gray!20} Electrode  I & 1.8 & 0.04 & 1.46 & 0.203 & 1.1e-73  \\
 Electrode  J & 1.77 & 0.04 & 1.50 & 0.156 & 2.4e-79  \\
\rowcolor{gray!20}Electrode  K & 1.74 & 0.056 & 1.43 & 0.172 & 4.4e-82  \\
 Electrode  L & 1.78 & 0.033 & 1.42 & 0.186 & 2.7e-95  \\
\rowcolor{gray!20}Electrode  M & 1.70 & 0.07 & 1.35 & 0.154 & 4.6e-92  \\
Electrode  N & 1.64 & 0.096 & 1.51 & 0.161 & 1.5e-27  \\
\rowcolor{gray!20}Electrode  O & 1.54 & 0.149 & 1.50 & 0.160 & 0.003  \\
Electrode  P & 1.74 & 0.063 & 1.52 & 0.157 & 7.4e-71  \\
\rowcolor{gray!20}Electrode  Q & 1.58 & 0.089 & 1.33 & 0.129 & 1.3e-81  \\
 Electrode  R & 1.61 & 0.084 & 1.31 & 0.115 & 8.2e-96  \\
 \rowcolor{gray!20}Electrode  S & 1.63 & 0.108 & 1.43 & 0.136 & 1.3e-61  \\
Electrode  T & 1.61 & 0.094 & 1.36 & 0.126 & 3.4e-81  \\
\rowcolor{gray!20}Electrode  U & 1.65 & 0.076 & 1.39 & 0.148 & 3.4e-79  \\
Electrode  V & 1.72 & 0.052 & 1.49 & 0.157 & 8.7e-75  \\
\hline
\end{tabular}
\hypertarget{TABLE2}{}
\end{table}
\begin{multicols}{2}


\end{multicols}
\begin{figure*}[!h]
  \includegraphics[width=\textwidth]
  {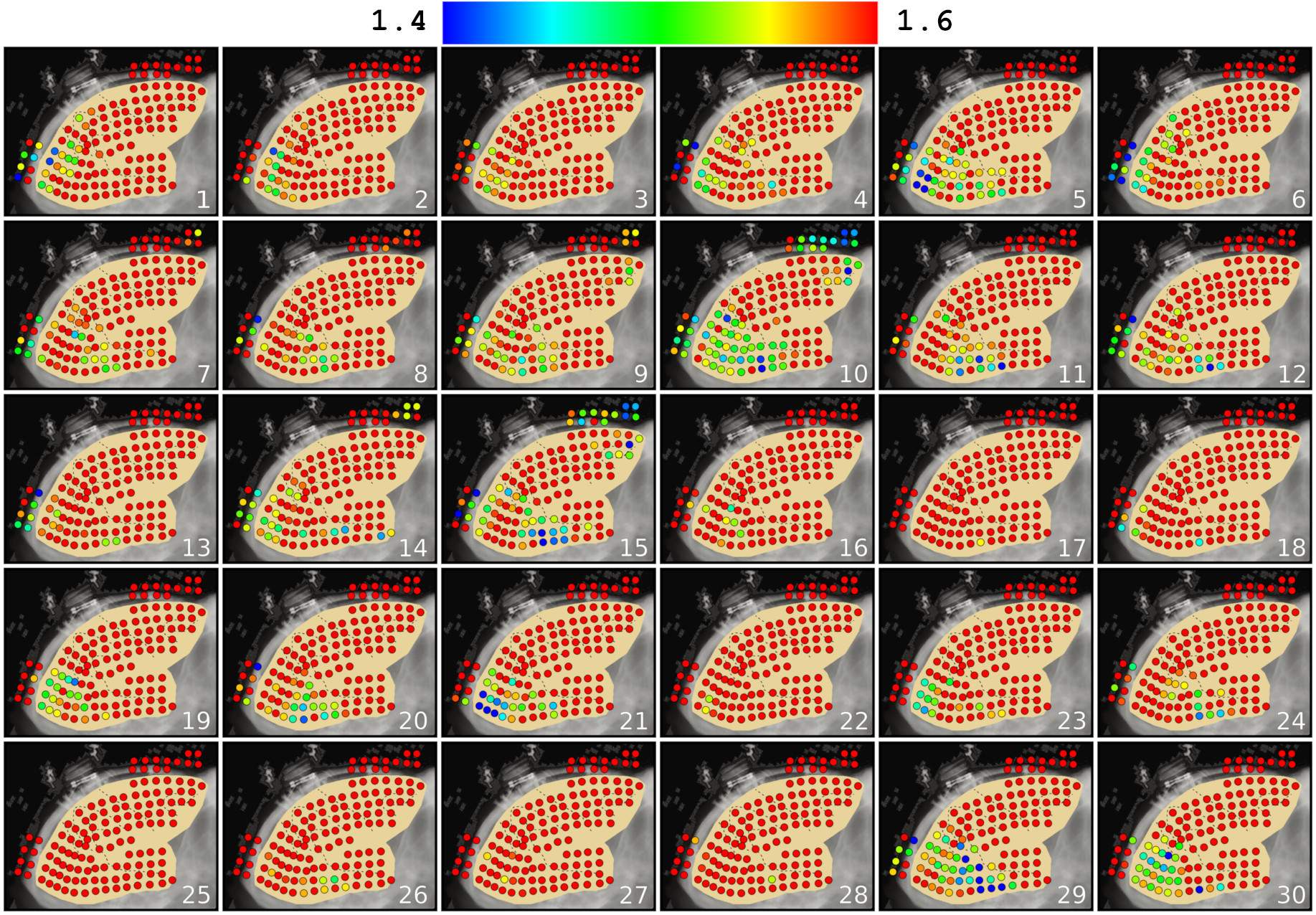}
     \caption{\textbf{Fractal Dimension during awake resting-state conditions}. This Figure shows the Fractal Dimension values over the locations of their respective electrodes. Sub-Figures were estimated sequentially throughout time at every 2.5 seconds, being this the time interval between each Sub-Figure and its subsequent one. The color gradient indicates the magnitude of the Fractal Dimension. In this Figure, it is possible to visualize the characteristic patterns of the complexity in awakened resting-state conditions. It is noticeable that under these conditions, most of the electrodes present Fractal Dimension values of around 1.6 and over. In addition, it is possible to observe that electrodes located at the occipital regions showed some tendency to display variation (see $Sub$-$Figures$ 5, 10, 11, 12, 19, 29, and 30).
}
     
     \hypertarget{FIGURE8}{}
 \end{figure*}
 \begin{multicols}{2}


\end{multicols}
\begin{figure*}[!h]
  \includegraphics[width=\textwidth]
  {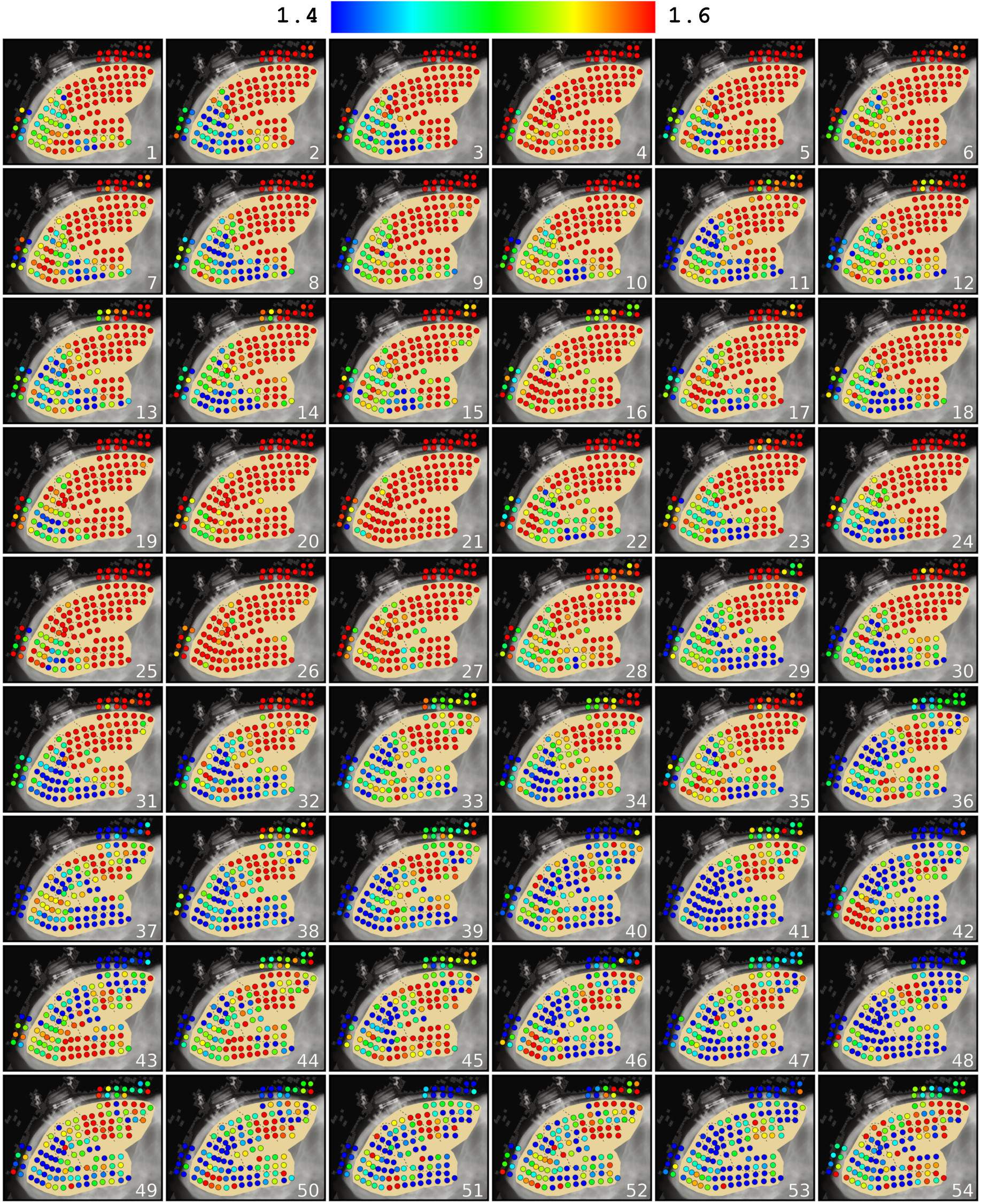}
  \captionsetup{format=plain}
     \caption{\textbf{Fractal Dimension during the transition to Ketamine-Medetomidine induced unconsciousness.} 
This Figure shows the Fractal Dimension values over the locations of their respective electrodes. Sub-Figures were estimated sequentially throughout time at every 2.5 seconds, being this the time interval between each Sub-Figure and its subsequent one. The color gradient indicates the magnitude of the Fractal Dimension.
In this Figure, it is possible to visualize the first alterations that happened in the patterns of the awake resting-state condition following the administration of the anesthetics. Primarily some regions of the temporal and occipital lobes presented a reduction, while frontal electrodes did not display expressive alterations (see $Sub$-$Figures$ 29, 30, 31,  and 32). Thereafter, a reduction in the electrodes of the frontal lobe was verified (see $Sub$-$Figures$ 37, 40, 42, 43, 46, 48, 50, 51, and 53). A remarkable characteristic observed during these first moments of the transition was that the central sulcus and nearby areas showed a tendency to exhibit high Fractal Dimension (see $Sub$-$Figures$ 33, 34, 36, 37, 38, 39, 41, 44, 46, 47, 48, 49, 50, 51, and 52). $Sub$-$Figures$ 26 and 27 had patterns characteristic of awake conditions, whereas $Sub$-$Figures$ 39 and 40 already displayed substantially different features. Considering that the time interval among consecutive Sub-Figures is 2.5 seconds, we infer that the first abrupt changes in the complexity of cortical activity as a whole took place within about 30 to 40 seconds.
}
    \hypertarget{FIGURE9}{} 
 \end{figure*}
 \begin{multicols}{2}


\end{multicols}
\begin{figure*}[!h]
  \includegraphics[width=\textwidth]
  {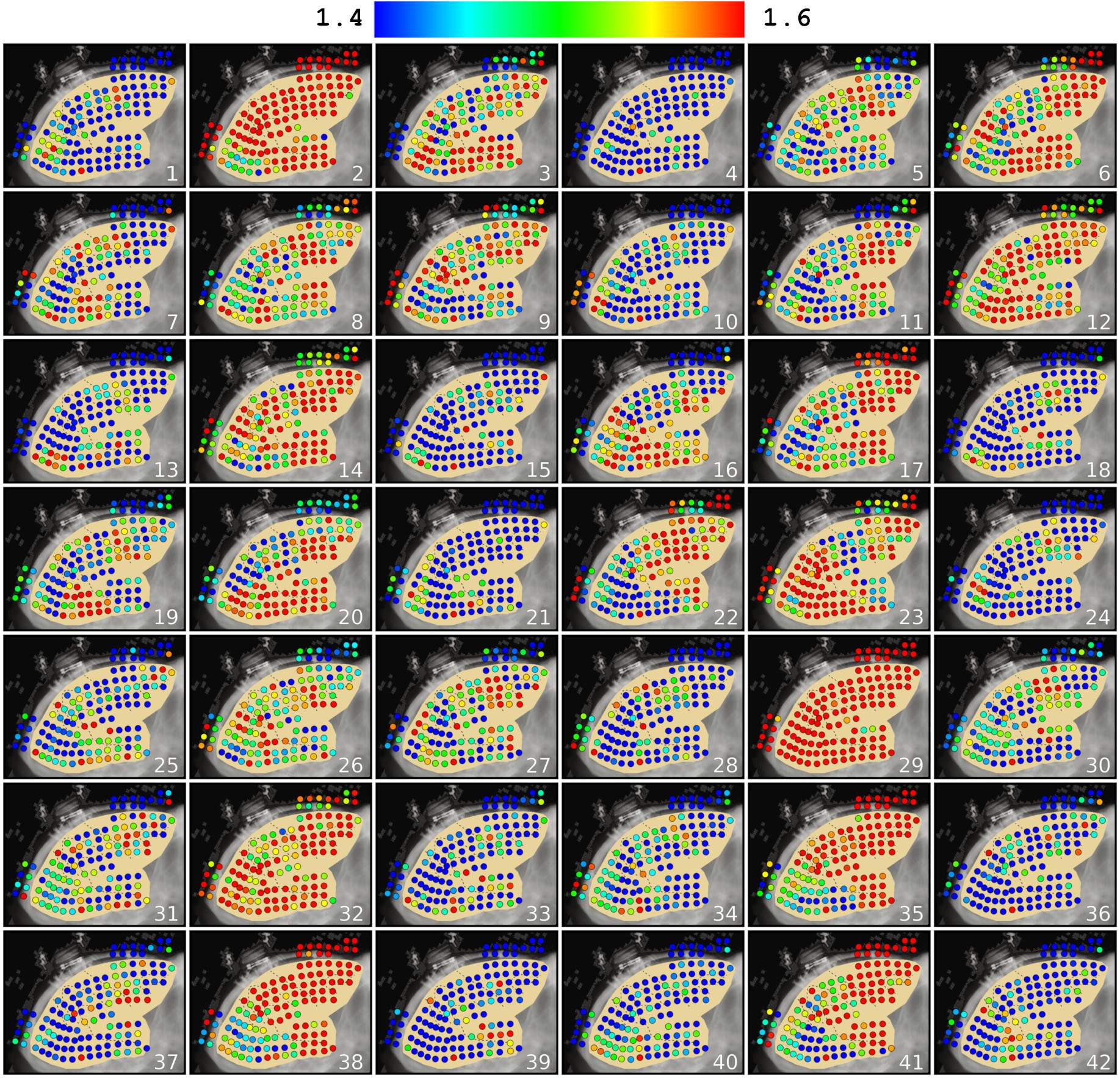}
     \caption{ \textbf{Fractal Dimension during general anesthesia.} This Figure shows the Fractal Dimension values over the locations of their respective electrodes. Sub-Figures were estimated sequentially throughout time at every 2.5 seconds, being this the time interval between each Sub-Figure and its subsequent one. The color gradient indicates the magnitude of the Fractal Dimension.
It is possible to visualize the characteristic patterns as well as the dynamics of the Fractal Dimension during the state of general anesthesia; it is noticeable that the cortex alternates between high complexity states that appear similar to those observed during alert conditions (see $Sub$-$Figures$ 2, 6, 14, 17, 23, 29, 32, 35, 38, and 41 ) and low complexity states in which the vast majority of cortical areas display reduced Fractal Dimension values (see $Sub$-$Figures$ 1, 4, 10, 13, 15, 18, 19, 21, 24, 25, 27, 28, 30, 31, 33, 34, 36, 37, 39, 40, and 42 ). It was also observed that even at periods of low complexity, there was still a tendency for the occipital-frontal and central sulcus to assume slightly higher values than the rest of the cortex (see $Sub$-$Figures$ 5, 8, 9, 11, 12, 17, 19, 20, 22, 26, and 27). That was a recurrent pattern that has shown up with a relative frequency over general anesthesia conditions, although not always present throughout the entire time.
 }
     \hypertarget{FIGURE10}{}
 \end{figure*}
 \begin{multicols}{2}


\end{multicols}
\begin{figure}[h]
\begin{subfigure}{.5\textwidth}
  \centering
  \includegraphics[width=1\linewidth]{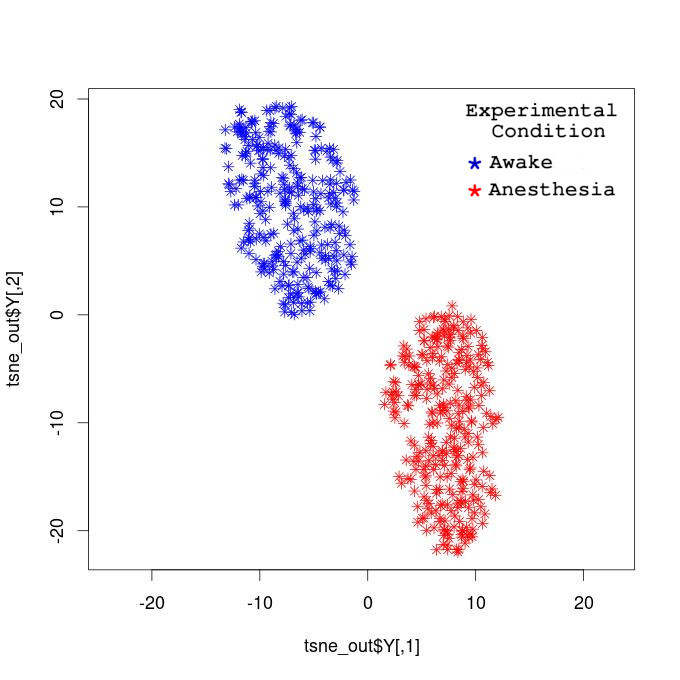}
  \caption{Pemutation Entropy}
  \label{fig11:sfigA}
\end{subfigure}%
\begin{subfigure}{.5\textwidth}
  \centering
  \includegraphics[width=1\linewidth]{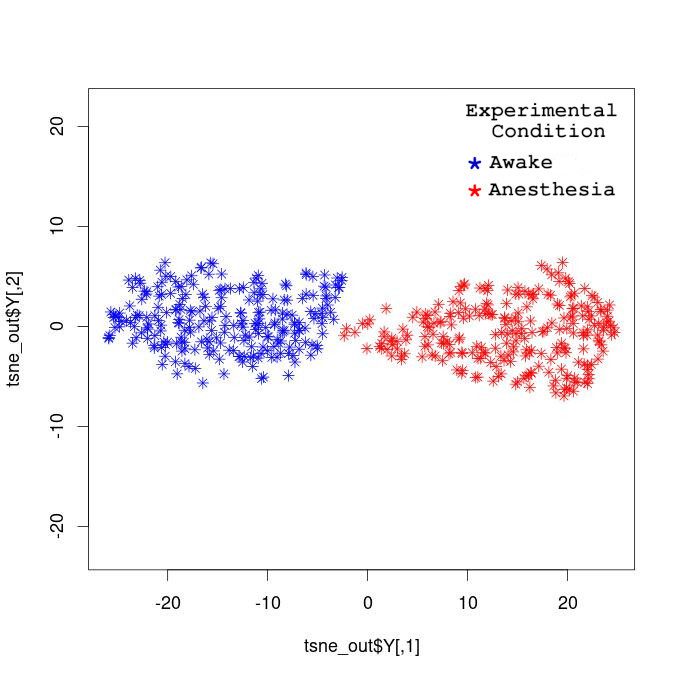}
  \caption{Fractal Dimension}
  \label{fig11:sfigB}
\end{subfigure}%
\caption{\textbf{Projections of the time-resolved vector states onto a 2-D map.} Projections of the time-resolved vector states (128-D) respective to the Permutation Entropy (Sub-Figure A) and the Fractal Dimension (Sub-Figure B) onto a bi-dimensional map by the use of the t-SNE algorithm. Alert resting-state conditions are indicated in blue, and general anesthesia in red; see legends. By examining this Figure, it is verified that in both methods, the Permutation Entropy and the Fractal Dimension, the projections of the alert state (indicated in blue) and general anesthesia (indicated in red) occupy distinct regions of the map. This result thus evidences that concerning the complexity of the cortical electrophysiological activity, alert resting-state and general anesthesia constitute two different states distinguishable at every moment.
}
\label{figure11}
\hypertarget{FIGURE11}{}
\end{figure}

\clearpage
\begin{multicols}{2}

\bibliography{bibfile2}

\begin{thebibliography}{5}
\providecommand{\natexlab}[1]{#1}
\providecommand{\url}[1]{\texttt{#1}}
\expandafter\ifx\csname urlstyle\endcsname\relax
  \providecommand{\doi}[1]{doi: #1}\else
  \providecommand{\doi}{doi: \begingroup \urlstyle{rm}\Url}\fi

\bibitem[Bandt and Pompe(2002)]{bandt2002permutation}
Christoph Bandt and Bernd Pompe.
\newblock Permutation entropy: a natural complexity measure for time series.
\newblock \emph{Physical review letters}, 88\penalty0 (17):\penalty0 174102,
  2002.

\bibitem[Higuchi(1988)]{higuchi1988approach}
Tomoyuki Higuchi.
\newblock Approach to an irregular time series on the basis of the fractal
  theory.
\newblock \emph{Physica D: Nonlinear Phenomena}, 31\penalty0 (2):\penalty0
  277--283, 1988.

\bibitem[Krijthe et~al.(2018)Krijthe, van~der Maaten, and
  Krijthe]{krijthe2018package}
Jesse Krijthe, Laurens van~der Maaten, and Maintainer~Jesse Krijthe.
\newblock Package ‘rtsne’, 2018.

\bibitem[Nagasaka et~al.(2011)Nagasaka, Shimoda, and
  Fujii]{nagasaka2011multidimensional}
Yasuo Nagasaka, Kentaro Shimoda, and Naotaka Fujii.
\newblock Multidimensional recording (mdr) and data sharing: an ecological open
  research and educational platform for neuroscience.
\newblock \emph{PloS one}, 6\penalty0 (7):\penalty0 e22561, 2011.

\bibitem[Van~der Maaten and Hinton(2008)]{van2008visualizing}
Laurens Van~der Maaten and Geoffrey Hinton.
\newblock Visualizing data using t-sne.
\newblock \emph{Journal of machine learning research}, 9\penalty0 (11), 2008.

\end{thebibliography}

\renewcommand{\figurename}{Supplementary Figure}
\setcounter{figure}{0}
\begin{wrapfigure}{l}{0.6\textwidth}

  \includegraphics[width=0.99\linewidth]
  {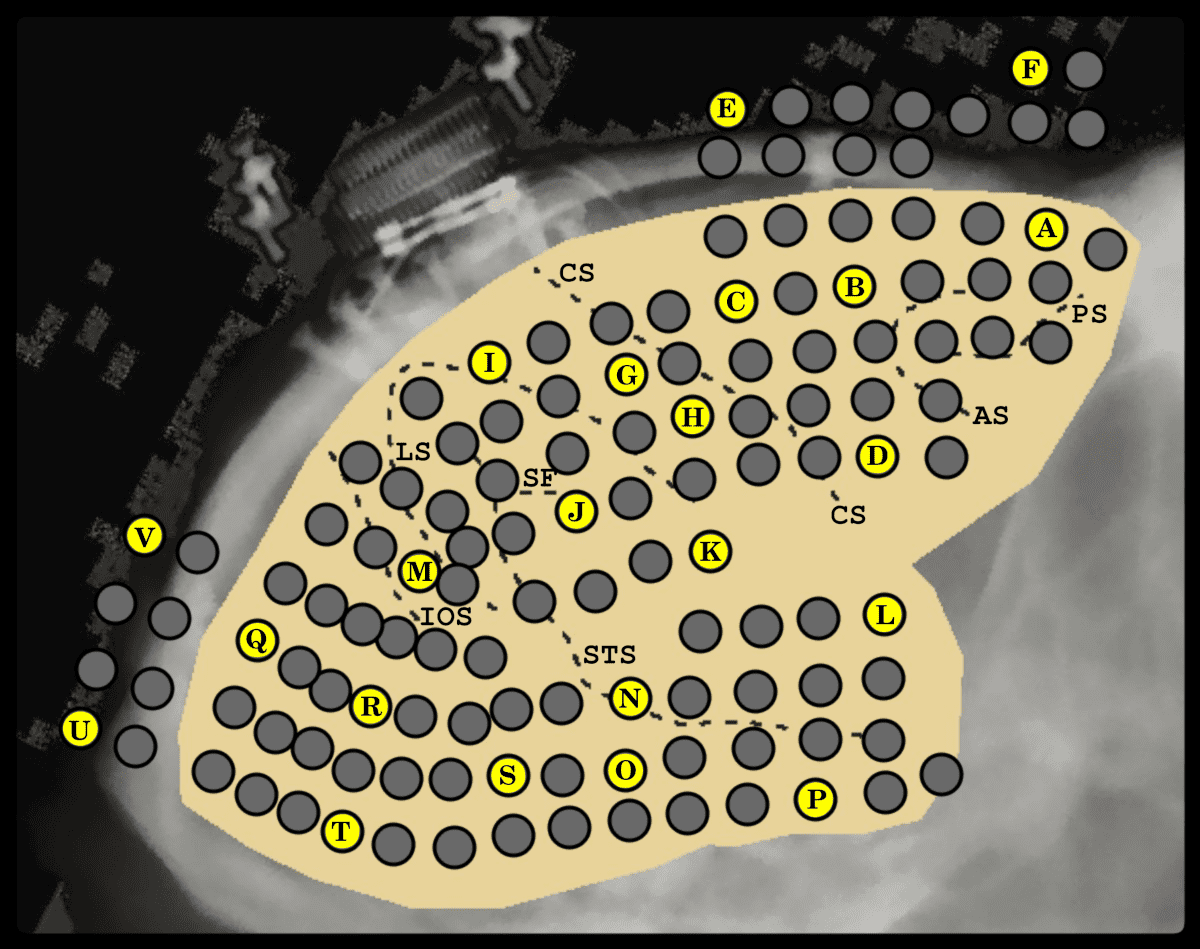}
  \captionsetup{format=plain}
       \caption{\textbf{Electrodes position reference}. This figure shows the position of the electrodes according to their location over the cortical surface. The correspondence of the electrodes referred on the \protect\hyperlink{FIGURE1}{$Figure \cdot 1$}, \protect\hyperlink{FIGURE2}{$Figure \cdot 2$}, \protect\hyperlink{FIGURE6}{$Figure \cdot 6$}, \protect\hyperlink{FIGURE7}{$Figure \cdot 7$}, \protect\hyperlink{TABLE1}{$Table \cdot 1$}, and \protect\hyperlink{TABLE2}{$Table \cdot 2$}     can be checked according to the corresponding letters from  A to V.  Electrodes E and F are placed on the medial frontal walls. Electrodes U and V are placed on the medial occipital walls. CS: central sulcus; PS: principal sulcus; AS: arcuate sulcus; IOS: inferior occipital sulcus; SF: sylvian fissure; IPS: intra parietal sulcus; LS: lunate sulcus; STS: superior temporal sulcus. Figure adapted from: (\texttt{http://neurotycho.org}
).}
   \hypertarget{FIGURE12}{}  
 \end{wrapfigure}
\end{multicols}

\end{document}